\documentclass{emulateapj}
\usepackage{graphicx}
\newcommand{\secpoint}{\mbox{$''\mskip-7.6mu.\,$}}

\slugcomment{Received 06/23/16; accepted 09/20/16}

\begin{document}

\title{Mid-Infrared Colors of Dwarf Galaxies: Young Starbursts Mimicking Active Galactic Nuclei}

\shorttitle{Mid-Infrared Colors of Dwarf Galaxies}
\shortauthors{HAINLINE ET AL.}

\author{\sc Kevin N. Hainline}
\affil{Steward Observatory, University of Arizona, 933 North Cherry Avenue, Tucson, AZ 85721, USA}

\author{Amy E. Reines\altaffilmark{1}}
\affil{National Optical Astronomy Observatory, Tucson, AZ 85726, USA}

\author{Jenny E. Greene}
\affil{Department Astrophysical Sciences, Princeton University, Princeton, NJ 08544}

\author{Daniel Stern}
\affil{Jet Propulsion Laboratory, California Institute of Technology, 4800 Oak Grove Drive, Mail Stop 169-221, Pasadena, CA 91109, USA}

\altaffiltext{1}{Hubble Fellow}

\begin{abstract}

Searching for active galactic nuclei (AGN) in dwarf galaxies is important for our understanding of the seed black holes that formed in the early Universe.  Here, we test infrared selection methods for AGN activity at low galaxy masses.  Our parent sample consists of ~18,000 nearby dwarf galaxies ($\mathrm{M}_{*} < 3\times10^9 \;\mathrm{M}_{\sun}$, $z<0.055$) in the Sloan Digital Sky Survey with significant detections in the first three bands of the AllWISE data release from the Wide-field Infrared Survey Explorer (\textit{WISE}).  First, we demonstrate that the majority of optically-selected AGNs in dwarf galaxies are not selected as AGNs using \textit{WISE} infrared color diagnostics and that the infrared emission is dominated by the host galaxies.  We then investigate the infrared properties of optically-selected star-forming dwarf galaxies, finding that the galaxies with the reddest infrared colors are the most compact, with blue optical colors, young stellar ages and large specific star formation rates.  These results indicate that great care must be taken when selecting AGNs in dwarf galaxies using infrared colors, as star-forming dwarf galaxies are capable of heating dust in such a way that mimics the infrared colors of more luminous AGNs.  In particular, a simple $\mathrm{W1}-\mathrm{W2}$ color cut alone should not be used to select AGNs in dwarf galaxies.  With these complications in mind, we present a sample of 41 dwarf galaxies worthy of follow-up observations that fall in \textit{WISE} infrared color space typically occupied by more luminous AGNs.

\end{abstract}

\keywords{cosmology: observations -- galaxies: evolution -- galaxies: dwarf galaxies -- galaxies: active galactic nuclei}

\section{Introduction}
\label{sec:intro}

There is now an overwhelming body of evidence suggesting that all massive galaxies host a central supermassive black hole which grows alongside the stellar population \citep{kormendy1995,kormendy2013}. In addition, a relationship has been observed between the central black hole mass and the galaxy bulge stellar velocity dispersion that spans many orders of magnitude \citep{gebhardt2000, ferrarese2000}. The origin of this relationship is still not well understood, but current theories imply that galaxy mergers and interactions play a role, both in increasing the stellar mass of a galaxy and in driving gas towards the centers of galaxies, feeding black holes \citep{hopkins2008, koss2010, ellison2011, bessiere2012, sabater2013}. However, this view of galaxy growth implies the existence of low mass ``seed'' black holes that must have existed at high redshift. To understand these difficult-to-observe objects, researchers have turned to observations of nearby dwarf galaxies which may host analogous lower-mass black holes (for a review, see Reines \& Comastri 2016, submitted). By assembling large samples of low-mass black holes, it may be possible to distinguish between the different proposed theoretical scenarios for their creation: these objects may be remnants of massive Population III stars \citep{bromm2011}, a result of direct collapse of primordial dense gas \citep{haehnelt1993, lodato2006, begelman2006, vanwassenhove2010}, or perhaps they are the end product of very massive stars formed through stellar mergers in dense star clusters \citep{gurkan2004, freitag2006,goswami2012, giersz2015, lutzgendorf2016}.

Assembling these large samples of low-mass black holes is made difficult by the fact that resolving their gravitational sphere of influence is currently not feasible at distances larger than a few Mpc. However, active galactic nuclei (AGN) emission across the electromagnetic spectrum can be used to infer the existence of a black hole. Observations at optical wavelengths have been used to uncover AGNs in NGC 4395 \citep{filippenko1989, filippenko2003} and POX 52 \citep{barth2004}, while data at X-ray and radio wavelengths have been used to find AGNs in both Henize 2-10 \citep{reines2011, reines2012} and the dwarf galaxy pair Mrk 709 \citep{reines2014}. Larger samples of low-mass AGNs have been uncovered at optical \citep{greene2004, greene2007, dong2012, reines2013, moran2014} and X-ray \citep{lemons2015, mezcua2015, pardo2016} wavelengths, which have been targeted with successful follow-up observations \citep{baldassare2016}, including the discovery of a $5 \times 10^4 \; \mathrm{M}_\sun$ BH in RGG 118 \citep{baldassare2015}.

One important and often-used method for selecting luminous AGNs relies on observations made in the mid-IR, where dust, heated by the central accreting black hole, reprocesses the light and emits with a characteristic red IR power-law spectrum. Infrared emission only minimally suffers from nuclear and galaxy-scale obscuration, and so mid-IR observations have successfully uncovered large numbers of unobscured and obscured luminous AGNs and quasars \citep{lacy2004, lacy2013, stern2005, hickox2007, donley2008, ashby2009, assef2010, stern2012, mendez2013, hainline2014b}. The all-sky mid-IR coverage of the \textit{Wide-field Infrared Survey Explorer} \citep[\textit{WISE},][]{wright2010} has allowed for observations of large samples of objects, and multiple authors have proposed \textit{WISE} color schemes which select for the red AGN power law emission in the infrared \citep{jarrett2011, stern2012, mateos2012}. These selection methods rely on the fact that AGNs are capable of heating dust to temperatures well above what is observed from stellar processes in moderate to high-mass galaxies, and have demonstrated high levels of reliability when applied to these objects. 

Recently, \citet{satyapal2014} and \cite{sartori2015} used mid-IR selection methods to assemble large samples of low-mass galaxy AGN candidates from \textit{WISE} data. Under the assumption that \textit{WISE} selection targets optically obscured, ``hidden'' AGNs in these objects, these authors draw broad conclusions about the population of low-mass black holes. The \citet{satyapal2014} study targets ``bulgeless'' galaxies without optical evidence for an AGN, and the authors conclude that star formation is not the primary source of the IR emission in their sample. In addition, they propose that the fraction of galaxies hosting IR-selected AGN activity \textit{increases} at low masses. This puzzling trend was also seen (with lower significance) using a larger sample of dwarf galaxies by \cite{sartori2015}, who compared multiple AGN selection methods and concluded that dwarf galaxies with \textit{WISE} colors indicative of AGN activity are bluer and potentially may have more ongoing star formation and lower metallicities than those selected using optical emission lines. 

These results are intriguing in light of the fact that no other tracer of AGN activity has thus far uncovered such large samples of AGNs in dwarf galaxies. In most AGN selection regimes, more luminous AGNs are easier to find both because massive BHs have a higher Eddington limit and because of increasing confusion due to star formation at low mass. Thus, the observed increase in AGN fraction at the lowest dwarf galaxy masses is puzzling. Low-mass SMBHs that exist in dwarf galaxies power AGNs that have such low luminosities that star formation in their hosts becomes a significant source of contamination. In particular, it has been shown that low-metallicity dwarf starburst galaxies are capable of heating dust to very high temperatures \citep{hirashita2004, reines2008, izotov2011, izotov2014, griffith2011, remyruyer2015}, producing red mid-IR colors. This was recently explored in \citet{oconnor2016}, who found that galaxies with low stellar masses have predominantly red \textit{WISE} colors, which the authors associate with higher specific star formation rates (sSFR) in these galaxies. Thus, it may be that using common mid-IR AGN selection methods on dwarf galaxies results in the selection of a large number of star-forming galaxies that contaminate the samples, leading to erroneous conclusions about AGN fractions at these masses. 

In this paper, we use \textit{WISE} data to empirically examine the infrared colors of dwarf galaxies as a function of their properties as probed by Sloan Digital Sky Survey (SDSS) data. Our results suggest that star formation, and not AGN activity, is responsible for the red \textit{WISE} colors for the majority of the dwarf galaxy population \citep[also see][]{izotov2014}. We propose a small sample of dwarf galaxy AGN candidates that require follow-up observations to confirm black hole activity. 

In Section \ref{sec:sample}, we discuss our sample selection, and describe how these objects were matched to \textit{WISE} photometry. We start by exploring the infrared properties of dwarf galaxies which have optical spectroscopic evidence for an AGN in Section \ref{sec:IRpropertiesBPT}. We then investigate optically-selected star-forming galaxies and propose a sample of IR-selected AGN candidates in Section \ref{sec:IRproperties}. We expand our sample to include IR-selected AGN candidates without optical emission line flux measurements Section \ref{sec:noopticalemission}, after which we compare these objects to samples of IR-selected AGN candidates presented in the literature in Section \ref{sec:comparison}. Finally, we explore our \textit{WISE} detection limits in Section \ref{sec:detection} and we discuss our results and draw conclusions in Section \ref{sec:conclusions}. 

\section{Sample Selection and Data}
\label{sec:sample}

The dwarf galaxies we explore in this paper were selected from the NASA Sloan Atlas (NSA), a catalogue of galaxies at $z < 0.055$ selected from the Sloan Digital Sky Survey (SDSS) DR8 \citep{york2000,aihara2011}, where the observations were re-analyzed, resulting in improved photometry \citep{blanton2011}, and spectroscopy \citep{yan2011,yan2012}. In addition, stellar-masses for each object were derived with the \texttt{kcorrect} code of \cite{blanton2007}, which uses the stellar population synthesis models of \citet{bc2003} and the nebular emission-line models of \citet{kewley2001}. These masses are provided in the NSA in units of M$_{\sun}\;h^{-2}$, and in this paper we have assumed $h = 0.73$. 

Following \citet{reines2013}, we selected all galaxies with $\mathrm{M}_{*} < 3\times10^9 \;\mathrm{M}_{\sun}$, the approximate mass of the Large Magellanic Cloud \citep{vandermarel2002}. We matched those objects to the AllWISE data release \citep{cutri2013}, which provides increased sensitivity as compared to the previous \textit{WISE} All-Sky data release. We note that because of the polar orbit of the \textit{WISE} telescope, the total exposure time across the full sky is non-uniform and depends on both the sky position and the zodiacal foreground emission. As $\mathrm{W1}-\mathrm{W2}$ color forms the basis for both of the primary AGN selection criteria that we will explore in this study \citep[that of][]{jarrett2011, stern2012}, and given the importance of deep data at $\lambda < 5\mu$m for separating emission from dust heated by star formation and AGN activity, we chose to use the updated AllWISE data, with W1 (3.4 $\mu$m) and  W2 (4.6 $\mu$m) $5\sigma$ sensitivities of 54 $\mu$Jy and 71 $\mu$Jy, respectively (compared to 68 $\mu$Jy and 111 $\mu$Jy in the All-Sky data release). These sensitivities are the minimum depth of the survey estimated for low coverage sky away from the Galactic plane.

When cross-matching with the AllWISE catalogue, we used a $5''$ matching radius, and then chose the specific \textit{WISE} photometry for each object based on the \textit{WISE} ext\_flg value. Following the AllWISE Explanatory Supplement\footnote{http://wise2.ipac.caltech.edu/docs/release/allwise/expsup/}, for most objects (ext\_flg$ = 0$ or 4), we chose the ``profile-fitting'' photometric magnitudes, which are optimized for objects which are unresolved in \textit{WISE}. For those objects where ext\_flg$ = 1$,2, or 3, we used the ``standard'' aperture magnitudes, which better traces the photometry for objects that are resolved. Finally, for a small subset of objects with 2MASS photometry, where ext\_flg$ = 5$, we used the photometry using apertures scaled from the 2MASS XSC shape values. Each object was only included it in our sample if the signal-to-noise ratio (SNR) for the W1-, W2-, and W3-band photometry was greater than 3, yielding a sample of 18,482 dwarf galaxies with significant \textit{WISE} detections. The majority of the objects were removed from our final sample because of their low SNR W3 photometry  ($\sim58 \%$ of the total NSA sample, compared to $\sim4 \%$ and $\sim7 \%$ for W1 and W2, respectively). The median $\mathrm{W1}-\mathrm{W2}$ and $\mathrm{W2}-\mathrm{W3}$ color uncertainties are 0.07 and 0.19, respectively, which we show with error bars in of our \textit{WISE} color-color diagrams.

While we have attempted to account for resolution effects with our usage of the \textit{WISE} ext\_flg value to choose appropriate photometry for each object, we also wish to highlight some systematic effects that could affect our results. First, the W1, W2, and W3 bands have angular resolutions of $6\secpoint1$, $6\secpoint4$, and $6\secpoint5$, respectively \citep{wright2010}. As a result, for any object that is slightly resolved in W1 but not in W2, the resulting W1 flux would be underestimated, resulting in a redder $\mathrm{W1}-\mathrm{W2}$ color. This effect could potentially move objects into the \textit{WISE} color space spanned by AGNs. In addition, for objects with existing 2MASS photometry, which are largely well-resolved by \textit{WISE}, colors estimated using apertures scaled from the 2MASS XSC shape value might be artificially blue as these apertures may contain foreground stars. Furthermore, these apertures may not contain all of the infrared emission from the object, leading to a systematic underprediction of the fluxes \citep[see][for an examination of this effect]{cluver2014}. Because of these photometric effects, we have carefully visually examined the objects that have red \textit{WISE} colors indicative of potential AGN activity, although objects with ext\_flg$ = 5$ represent a small percentage of the total sample. 

\section{Mid-Infrared Properties of Dwarf Galaxies and AGN Candidates}
\label{sec:IRpropertiesDwarf}

\subsection{Optically-Selected AGNs and Composite Galaxies}
\label{sec:IRpropertiesBPT}

We begin our analysis of the IR properties of dwarf galaxies by exploring the IR colors of the AGNs and composite dwarf galaxies selected from \citet{reines2013}. These 136 dwarf galaxies were initially chosen from their optical spectroscopic properties and specifically their position on the BPT diagram \citep{baldwin1981}, a common optical emission-line diagnostic diagram that plots the flux ratio of [OIII]$\lambda$5007/H$\beta$ against [NII]$\lambda$6583/H$\alpha$\footnote{\citet{reines2013} also identified a sample of AGNs with broad H$\alpha$ emission as potential AGN candidates, but in this section we focus only on those objects in their sample with optical emission line ratios indicative of AGN activity.}. This diagram separates AGNs from star-forming galaxies based on commonly-used dividing lines, an empirical line given in \citet{kauffmann2003}, and a theoretical ``maximum starburst line'' from \citet{kewley2001}. Objects above the \citet{kewley2001} maximum starburst line on the diagram are often categorized as AGNs, while objects between the \citet{kauffmann2003} and \citet{kewley2001} lines are thought to have contributions to their emission line flux from both star formation and AGN activity, and are known as ``composite'' objects. 

	\begin{figure}[htbp]
	\epsscale{1.2} 
	\plotone{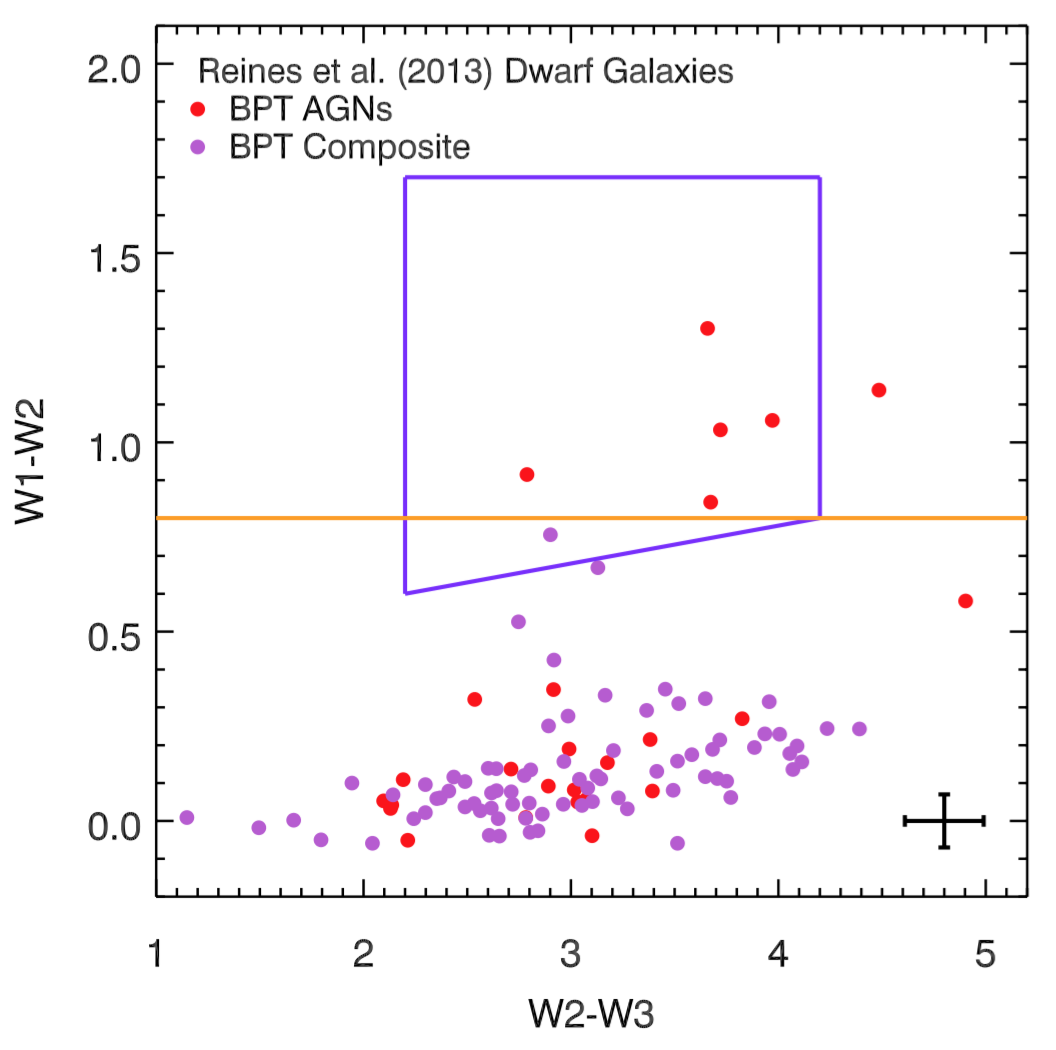}
	\caption{
	\label{fig:reinesagn} \textit{WISE} color-color diagram for the dwarf galaxies from \citet{reines2013} selected as AGN (red points) or composite galaxies (purple points) by their position on the BPT diagram. Four galaxies from the full \citeauthor{reines2013} sample did not have a match to the AllWISE catalogue, and 28 galaxies do not have significant detections in the W1, W2 or W3 bands, as required to place them on this plot. We plot two common AGN selection criteria with different colored lines. In blue, we plot the selection box from \citet{jarrett2011}, and in orange, we plot the $\mathrm{W1}-\mathrm{W2} > 0.8$ color criteria from \citet{stern2012}. We also plot the median $\mathrm{W1}-\mathrm{W2}$ and $\mathrm{W2}-\mathrm{W3}$ uncertainties with the error bars in the bottom-right corner.} 
	\epsscale{1.}
         \end{figure}

\begin{figure*}[htb]
	\centering
	  \begin{tabular}{@{}c@{}c@{}c@{}}
	   \includegraphics[width=.33\textwidth]{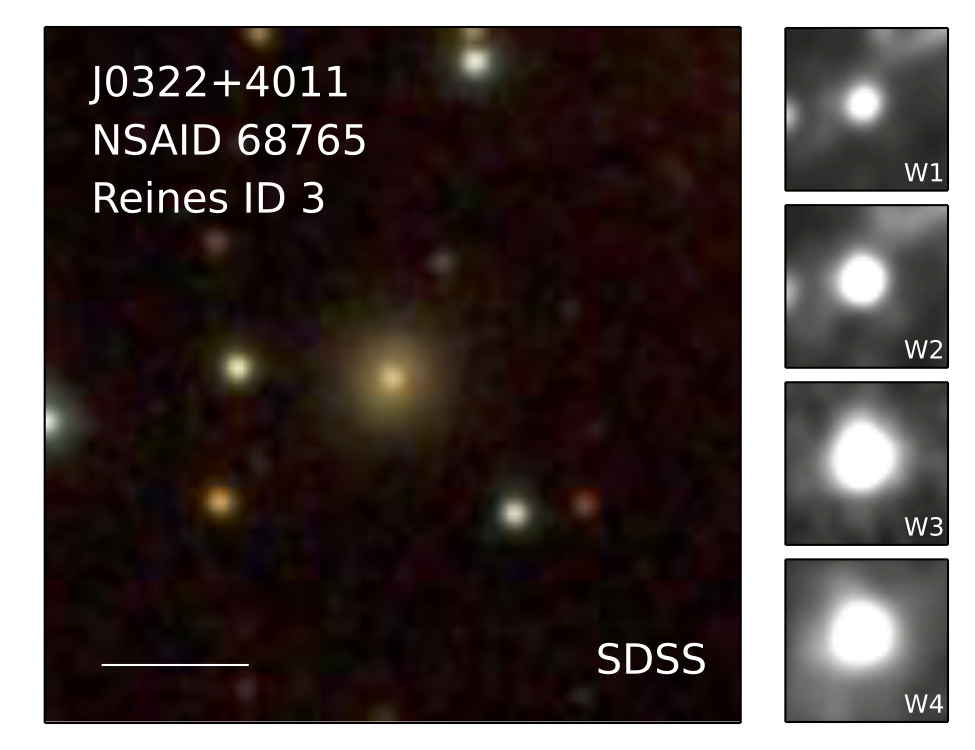}
	   \includegraphics[width=.33\textwidth]{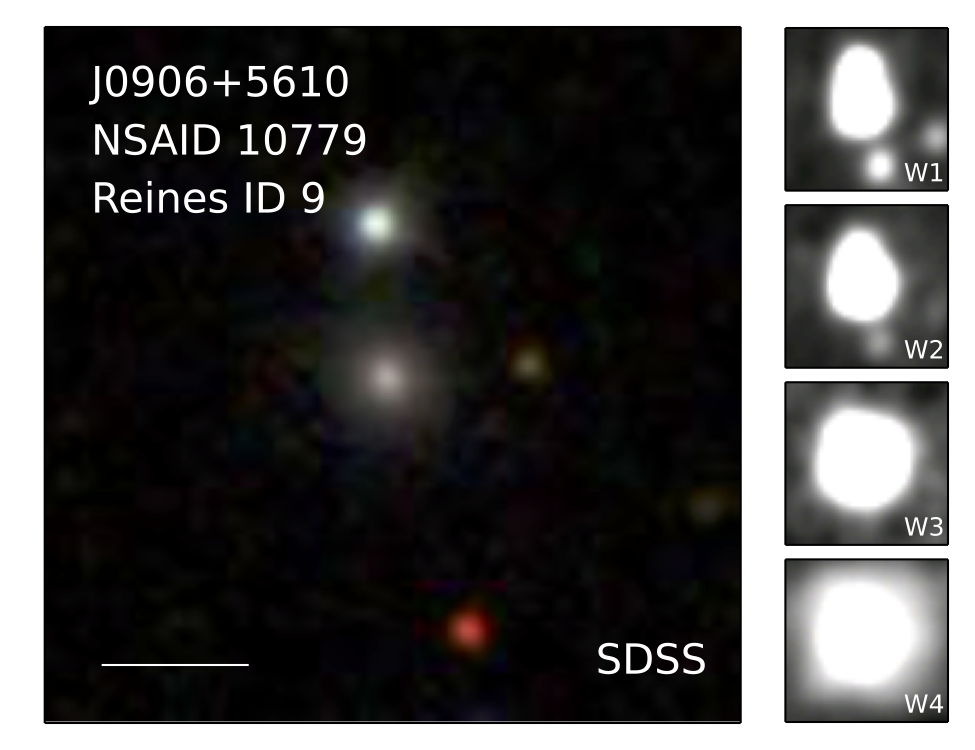}
	   \includegraphics[width=.33\textwidth]{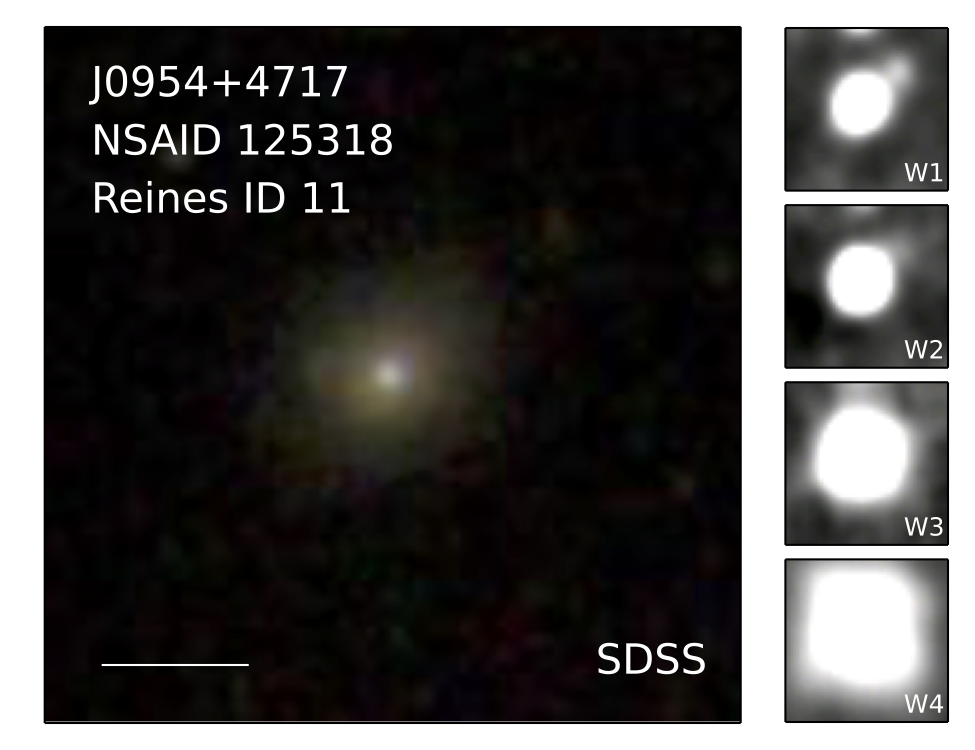} \\

	   \includegraphics[width=.33\textwidth]{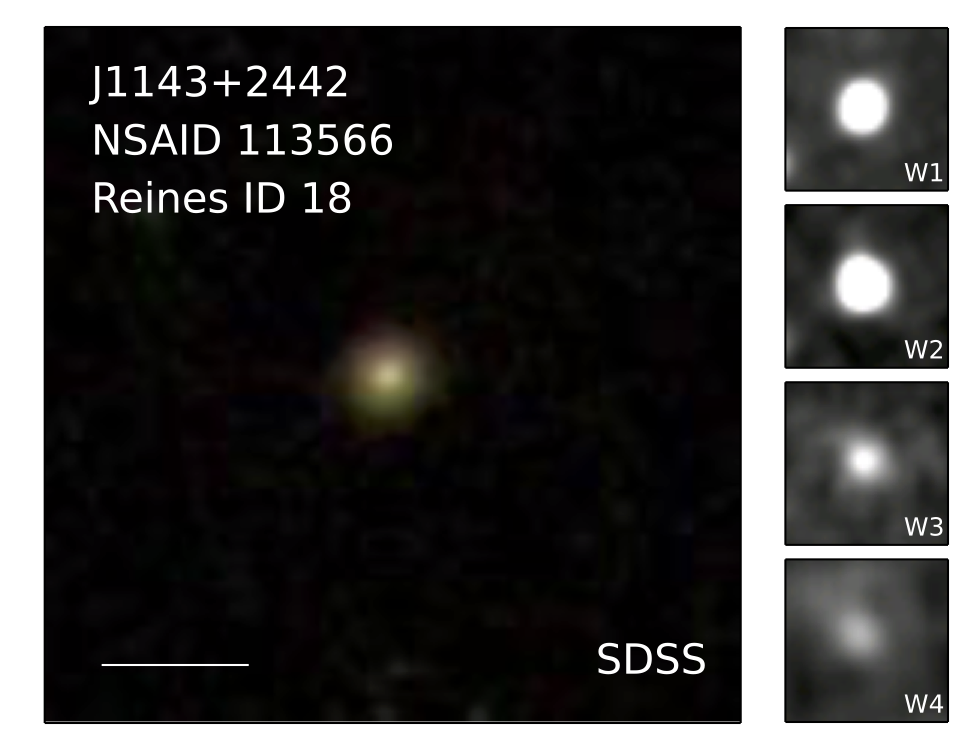}
	   \includegraphics[width=.33\textwidth]{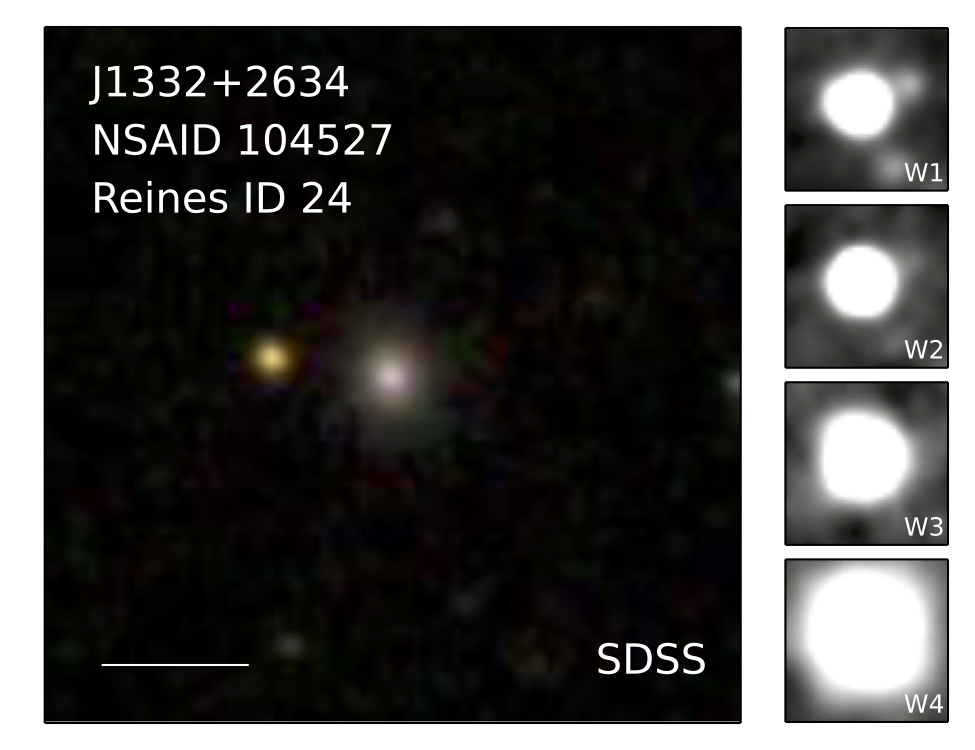}
	   \includegraphics[width=.33\textwidth]{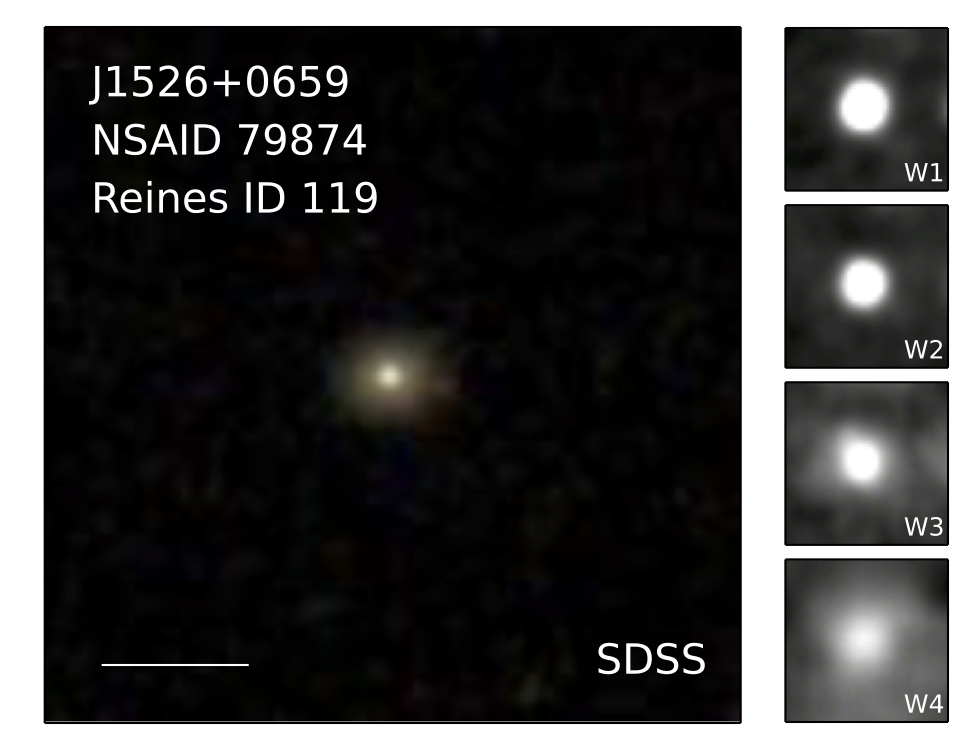} \\

	\end{tabular}
	\caption{
	\label{fig:bpt_agns_thumbnails} SDSS and \textit{WISE} thumbnails for the \citet{jarrett2011} \textit{WISE} AGN candidates in the BPT AGN and composite region of the BPT diagram \citep{reines2013}.  Each object is labelled with both the NSAID (see Table \ref{jarrettcandidates}) as well as the ID from \citet{reines2013} in the SDSS image. The SDSS multi-band image is on the left, and on the right, the \textit{WISE} W1, W2, W3, and W4 images are shown from top to bottom. In each panel, north is up, and east is to the left, and the bar in the bottom left of each SDSS panel is $10\arcsec$ in length. Both the SDSS and \textit{WISE} thumbnails are 48$\arcsec$ on a side. At the median redshift of our sample, $z = 0.03$, $1\arcsec = 0.6$ kpc.} 
         \end{figure*}

Of the 136 BPT AGNs and composites from \citet{reines2013}, 104 objects have significant \textit{WISE} detections as described in Section \ref{sec:sample} (four objects did not have a match to the AllWISE catalogue, and the other 28 objects do not have significant detections in the W1, W2 or W3 bands). We plot these 104 dwarf galaxies on the infrared AGN diagnostic \textit{WISE} color-color plot in Figure \ref{fig:reinesagn}, which compares the $\mathrm{W1}-\mathrm{W2}$ color to the $\mathrm{W2}-\mathrm{W3}$ color. We color the points based on whether they were classified as an optical AGN (red) or composite galaxy (purple). We also plot common infrared AGN selection criteria with colored lines. In blue, we plot the selection box from \citet{jarrett2011}, and in orange, we plot the $\mathrm{W1}-\mathrm{W2} > 0.8$ color criteria from \citet{stern2012}. These selection boxes select for luminous quasars by the presence of power-law emission that extends from the W1 through W3 bands due to reprocessed emission from the dust torus from around the accretion disk feeding the central supermassive black hole. As can be seen from Figure \ref{fig:reinesagn}, the majority of the dwarf galaxies selected as AGNs and composite objects by their optical emission would \textit{not} be selected as IR AGN on this diagram. Only six of the \citet{reines2013} BPT-AGN dwarf galaxies would be selected as AGN by the \citet{stern2012} criterion. Five BPT-AGN and one BPT-composite object would be selected as an AGN by the \citet{jarrett2011} criteria. We plot the SDSS and \textit{WISE} thumbnail images for these six objects in Figure \ref{fig:bpt_agns_thumbnails}. From this Figure, it can be seen that these optically-selected AGNs have significant \textit{WISE} detections out to the longest wavelengths probed by \textit{WISE}, evidence of an underlying infrared power-law from AGN-heated dust. As discussed in the introduction, more luminous AGNs are more likely to be found with \textit{WISE} selection, as their infrared emission can be observed above the emission from dust heated by host galaxy star formation. This helps to explain the general trends observed in Figure \ref{fig:reinesagn}, where five out of six of the objects in the \citet{jarrett2011} selection box are BPT-AGNs. 

The majority of both the BPT-AGNs and BPT-composite dwarf galaxies lie below the AGN selection regions in the area occupied by star-forming galaxies. These objects form a sequence that moves from blue $\mathrm{W1}-\mathrm{W2}$ and $\mathrm{W2}-\mathrm{W3}$ colors to the right and towards redder $\mathrm{W2}-\mathrm{W3}$ colors. For these objects, the AGN IR luminosity must be low enough that the reprocessed infrared emission is overwhelmed by the emission from both the galaxy's stellar continuum, the continuum emission from dust heated by star formation throughout host galaxy, and polycyclic aromatic hydrocarbon (PAH) emission. We will explore this in more detail in the next section. 

\subsection{Optically-Selected Star-Forming Dwarf Galaxies}
\label{sec:IRproperties}

	\begin{figure}[htbp]
	\epsscale{1.2} 
	\plotone{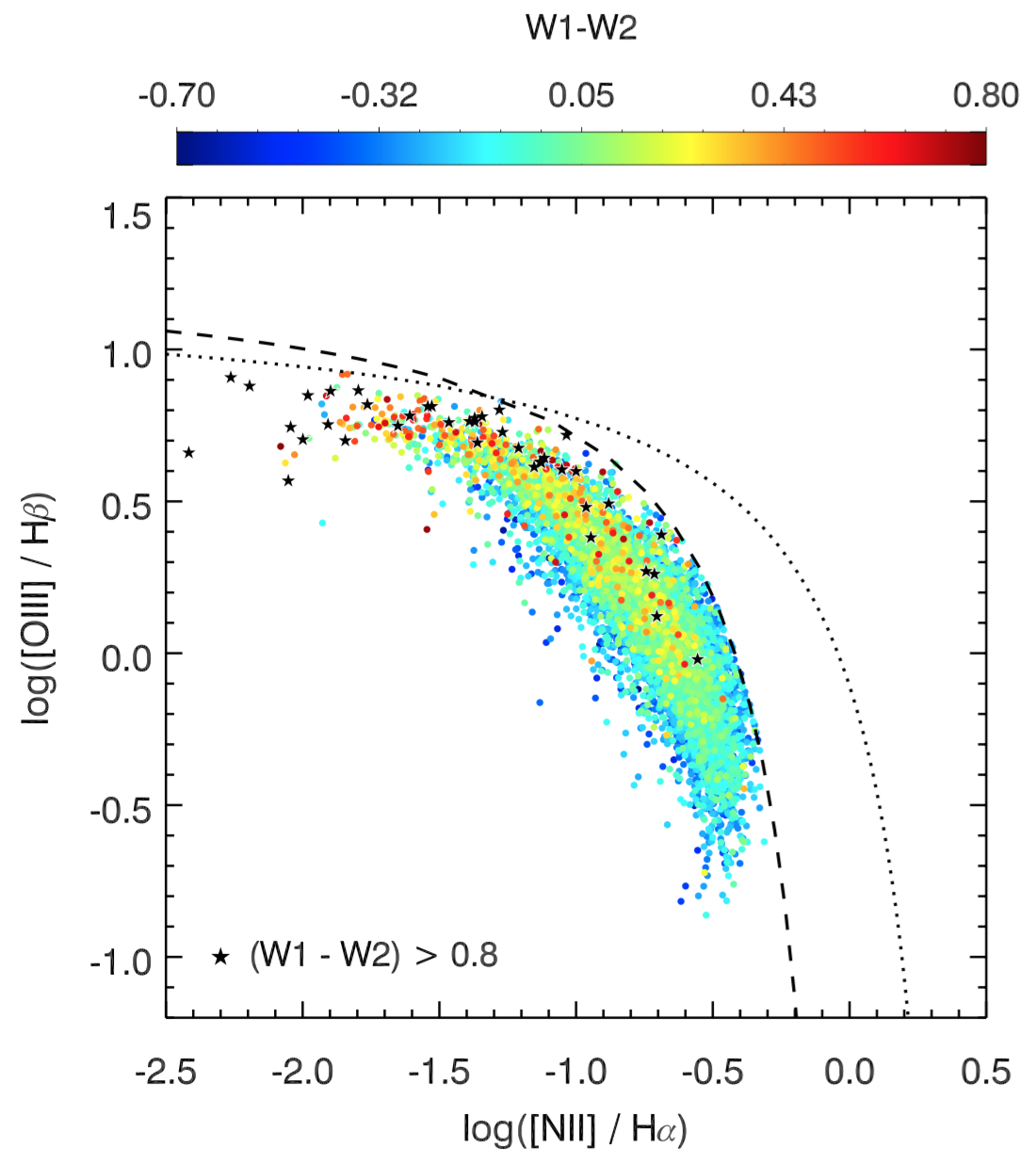}
	\caption{
	\label{fig:bptplot} BPT ionization diagram for the star-forming dwarf galaxies described in Section \ref{sec:IRproperties}. The points are colored by their W1-W2 infrared color, and those objects with $W1-W2 > 0.8$, the criteria used in \citet{stern2012} for selecting AGN candidates, are plotted with black stars. Overall, there is an observed trend in the figure between position to the top left of the star-forming sequence and red \textit{WISE} $W1-W2$ color.} 
	\epsscale{1.}
         \end{figure}

We now look at the dwarf galaxies in the NSA with line ratios dominated by star formation. We use a subsample (N = 14,013) of our \textit{WISE}-selected dwarf galaxies that have significant (SNR $> 3.0$) SDSS detections of the strong optical emission lines (H$\beta$, [OIII]$\lambda$5007, H$\alpha$ and [NII]$\lambda$6583) and fall in the star-forming part of the BPT diagram. We use the wealth of SDSS optical data to understand how dwarf galaxy SF properties relate to \textit{WISE} infrared colors and explore the potential origin of the red \textit{WISE} colors observed in some dwarf galaxies. In Figure \ref{fig:bptplot}, we show the star-forming subsample on the BPT diagram with points colored by $\mathrm{W1}-\mathrm{W2}$ infrared color. We plot those dwarf galaxies with $\mathrm{W1}-\mathrm{W2} > 0.8$, the selection criteria used to select AGNs from \citet{stern2012}, using black stars. For these star-forming dwarf galaxies, we can see that, generally, the $\mathrm{W1}-\mathrm{W2}$ color becomes redder as you move upwards and to the left on the sequence. Star-forming galaxies on the BPT diagram form a sequence in metallicity. Objects with the reddest \textit{WISE} colors are among the lowest metallicity galaxies in our sample and also have the largest ionization parameters \citep[see][for a review of how objects move on the star-forming region of the BPT diagram]{kewley2013}. 

\subsubsection{Physical Properties of Star-Forming Dwarf Galaxies}
\label{sec:derivingopticalproperties}

We can use the SDSS spectroscopic data to derive properties of the stellar populations in the dwarf galaxies for comparison to their \textit{WISE} colors. Some properties, such as galaxy size and optical color, are taken directly from the NSA, but in this section, we outline the methods we use to estimate the host galaxy star formation rate (SFR), stellar age, and ionizing luminosity.

\textit{Star Formation Rates.} The luminosity of H$\alpha$ can be used to measure the star formation rate for a galaxy, following \citet{kennicutt2012}:

\begin{equation} 
\log{(SFR [\mathrm{M}_{\sun} \; \mathrm{year}^{-1} ])} = \log{(L(\mathrm{H}\alpha)\; [\mathrm{erg} \; \mathrm{s}^{-1}])} - 41.27
\end{equation}

We have corrected the H$\alpha$ luminosity for the effects of extinction, using an intrinsic Balmer ratio of $F_{\mathrm{H}\alpha} / F_{\mathrm{H}\beta} = 2.86$, $R_V = 3.1$ and the \citet{odonnell1994} extinction curve. This method of estimating extinction is more valid in regions of low obscuration, and is not reliable for the most obscured regions. We also calculate the specific star formation rates (sSFR), the star formation rate divided by the stellar mass of a galaxy. Since the H$\alpha$ emission line fluxes were measured using $3''$ diameter fibers\footnote{The H$\alpha$ flux and EW estimates we use in this work are measured from SDSS fiber measurements and are dependent on the exact morphology and redshifts of the galaxies. As a result, we may be missing H$\alpha$ flux from star formation outside the fiber aperture. The majority of those objects with red \textit{WISE} colors are smaller than the SDSS aperture.}, we calculated the sSFR for each galaxy using masses estimated from the \textit{fiberflux} values given in the NSA. These fluxes, which were measured within a $3''$ aperture, were converted to a stellar mass using the method discussed in \citet{reines2015}, where we used the color-dependent mass-to-light ratio from \citet{zibetti2009}, adopting a solar absolute $i$-band magnitude of 4.56 mag \citep{bell2003}. While most of the sources have sSFR $< 1.0$ Gyr$^{-1}$, the majority of the red \textit{WISE} sources have sSFR $> 1.3$ Gyr$^{-1}$. In addition, we also use the Petrosian half-light radius from the NSA to estimate the SFR surface density ($\Sigma_{\mathrm{SFR}}$), the SFR divided by the area of the galaxy in kpc$^2$. 

\textit{Stellar Age.} The EW of the hydrogen Balmer lines anticorrelates with the age of a stellar population \citep{stasinska1996}. We estimate stellar ages by comparing measured H$\alpha$ values to STarburst 99 models \citep{leitherer1999}, assuming a \citet{salpeter1955} IMF with power-law $\alpha = 2.35$, and a metallicity of $Z = 0.004$ (where $Z = 0.02$ represents solar metallicity). The majority of the galaxies with red \textit{WISE} colors have high H$\alpha$ EWs and inferred stellar ages less than 7 Myr. 

\textit{Ionizing Luminosity.} We use the extinction-corrected H$\alpha$ luminosity to estimate the rate of ionizing photons ($Q_{\mathrm{Lyc}}$) for these objects following the procedure from \citet{reines2008}:

\begin{equation} \label{ionizingphotons}
\left( \frac{Q_{\mathrm{Lyc}}}{\mathrm{s}^{-1}} \right) \geq 2.25 \times 10^{12}  \left( \frac{T_e}{10^4 \mathrm{K}} \right)^{0.07} \left( \frac{L_{\mathrm{H}\alpha} / 2.86}{\mathrm{erg}\;\mathrm{s}^{-1}} \right)
\end{equation}

\noindent Here, we start from Equations (2) and (3) from \citet{condon1992}, and then assume Case B recombination, where $F_{\mathrm{H}\alpha} / F_{\mathrm{H}\beta} = 2.86$. We further assume an electron temperature of 10,000 K \citep{izotov1997}, and, as seen from the above equation, our estimated ionizing photon rate is only a lower limit due to the fraction of ionizing radiation that is absorbed by dust or escapes the star-forming regions before ionizing hydrogen atoms. The dwarf galaxies in our sample with the reddest $\mathrm{W1}-\mathrm{W2}$ colors have large H$\alpha$ luminosities and values of $Q_{\mathrm{Lyc}}$ equivalent to greater than several thousand O-type stars. 

\subsubsection{Mid-IR Colors of Star-Forming Dwarf Galaxies}
\label{sec:trends}

\begin{figure}[htbp]
	\epsscale{1.2} 
	\plotone{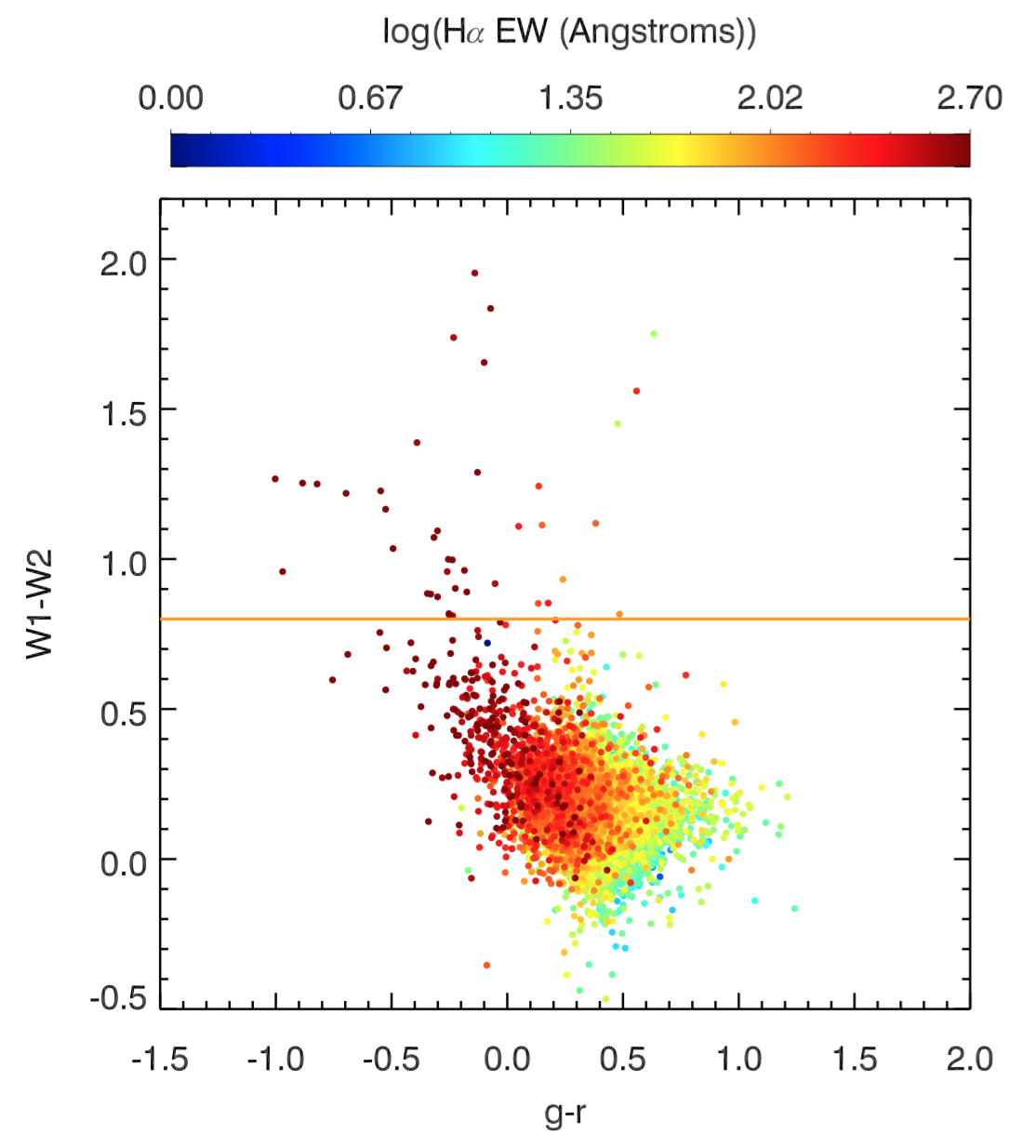}
	\caption{
	\label{fig:optical_infrared_color_plot} Optical to infrared color for the dwarf galaxies described in Section \ref{sec:IRproperties}. The host galaxies with the reddest \textit{WISE} colors are, on average, more optically blue, with the highest H$\alpha$ EWs and youngest ages.} 
	\epsscale{1.}
         \end{figure}

\begin{figure}[htbp]
	\epsscale{1.2}
	\plotone{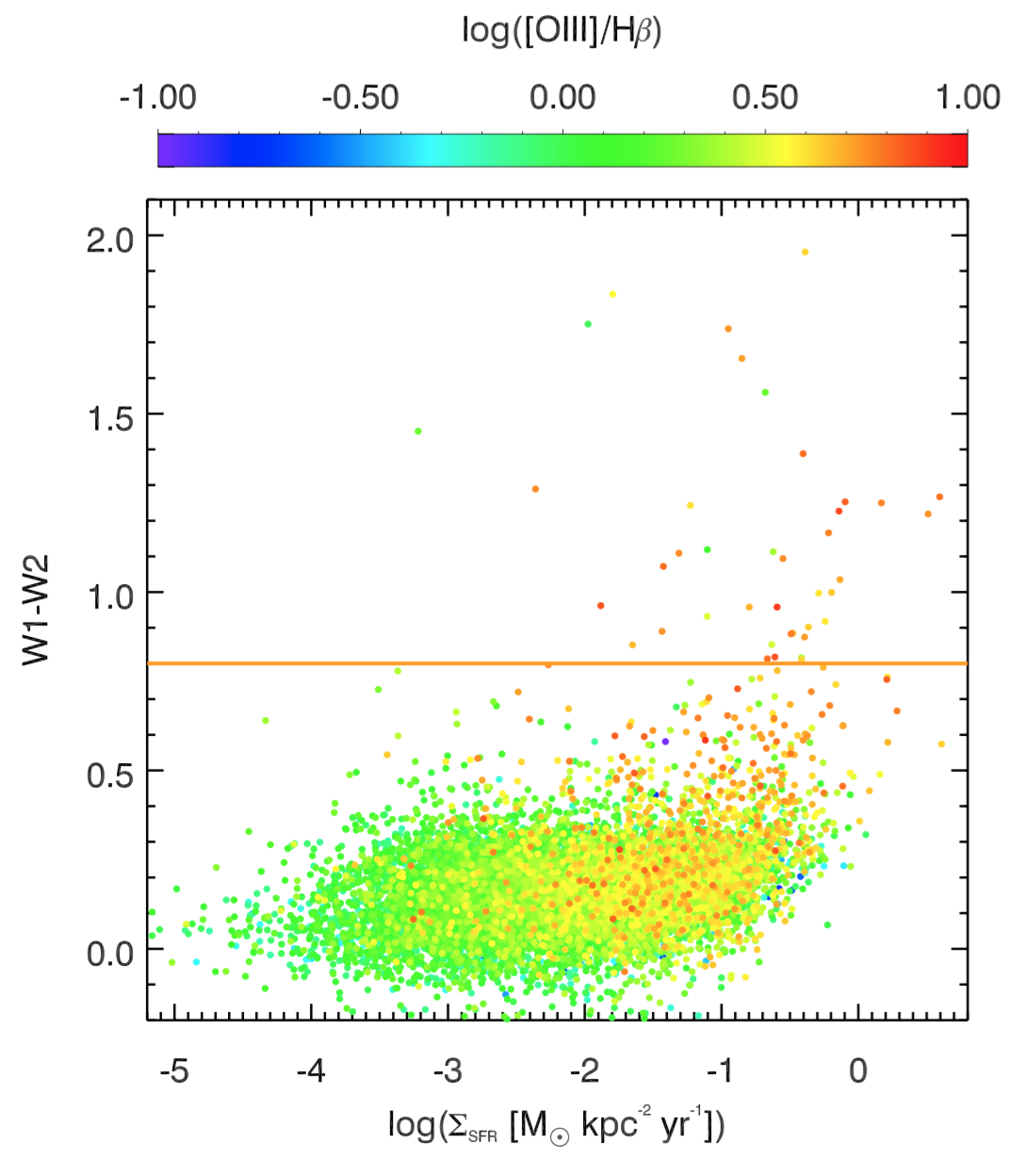}
	\caption{
	\label{fig:wisecolormorpho} \textit{WISE} $\mathrm{W1}-\mathrm{W2}$ color plotted against the SFR surface density ($\Sigma_{\mathrm{SFR}}$) for the dwarf galaxies described in Section \ref{sec:IRproperties}, with the points colored by [OIII]/H$\beta$ line ratio. Those galaxies with the reddest $\mathrm{W1}-\mathrm{W2}$ colors have the highest SFR surface densities.} 
	\epsscale{1.}
         \end{figure}

To investigate the origin of the infrared emission from the star-forming dwarfs, we plot the optical $g-r$ and infrared $\mathrm{W1}-\mathrm{W2}$ colors for these objects in Figure \ref{fig:optical_infrared_color_plot}. Here, we can see that the galaxies with the \textit{bluest} optical colors have the \textit{reddest} infrared colors. We have colored the points by the equivalent width (EW) of the H$\alpha$ emission line taken from the NSA, and we see that those dwarf galaxies with the reddest \textit{WISE} colors have the highest H$\alpha$ EW. We plot \textit{WISE} $\mathrm{W1}-\mathrm{W2}$ color as a function of $\Sigma_{\mathrm{SFR}}$ for each galaxy in Figure \ref{fig:wisecolormorpho}, and we see that the majority of the objects with the reddest $\mathrm{W1}-\mathrm{W2}$ colors also have the highest $\Sigma_{\mathrm{SFR}}$, further supporting the idea that we are observing compact young starbursts in these dwarf galaxies with red \textit{WISE} colors. We note that our usage of aperture fluxes to derive SFRs would lead to values of $\Sigma_{\mathrm{SFR}}$ for objects with Petrosian half-light radii larger than the SDSS fiber to be underestimated, although this effect does not significantly change the results seen in Figure \ref{fig:wisecolormorpho}.

\begin{figure*}[tb]
\centering
  \begin{tabular}{@{}cc@{}}
    \includegraphics[width=.5\textwidth]{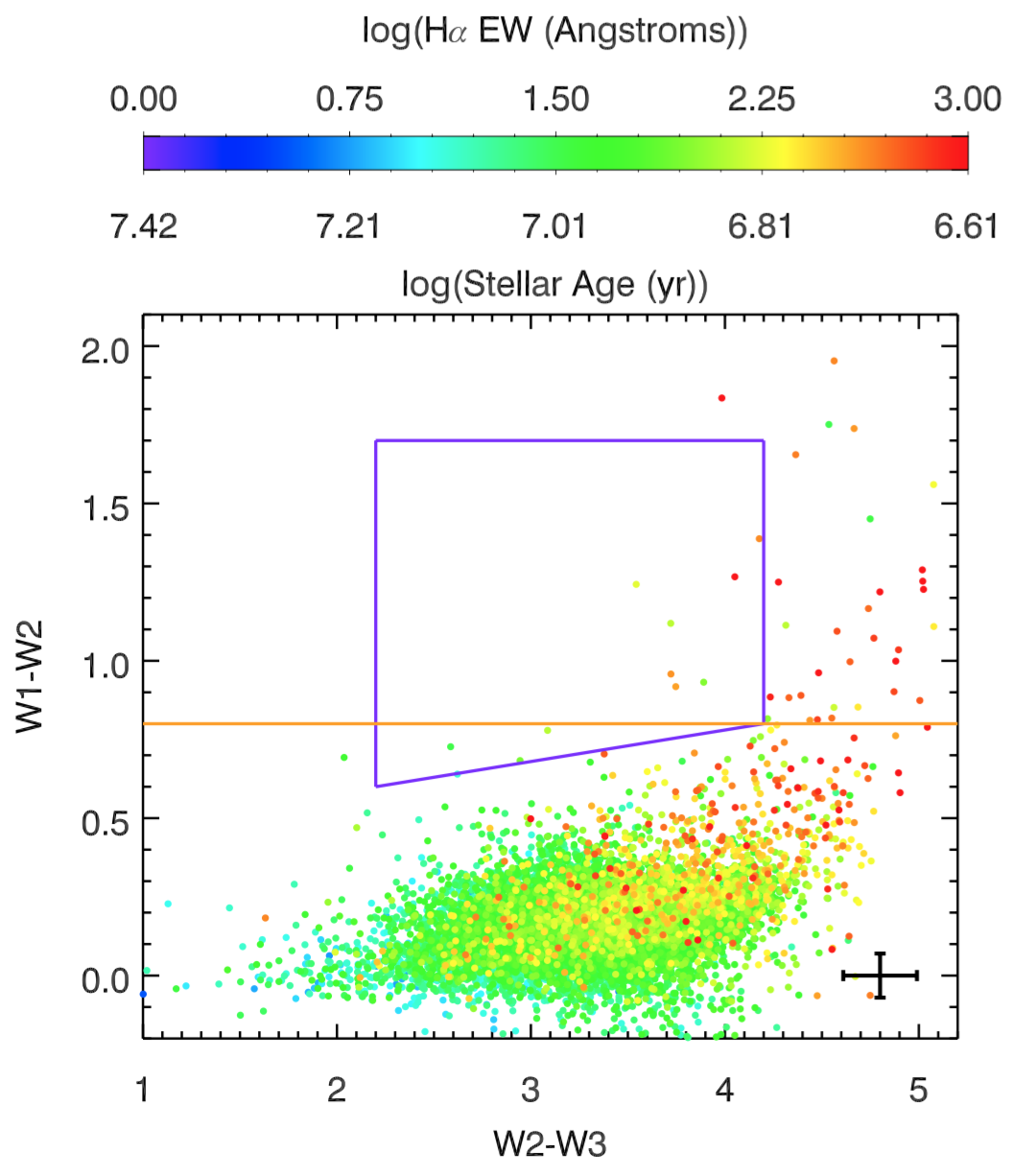} &
    \includegraphics[width=.5\textwidth]{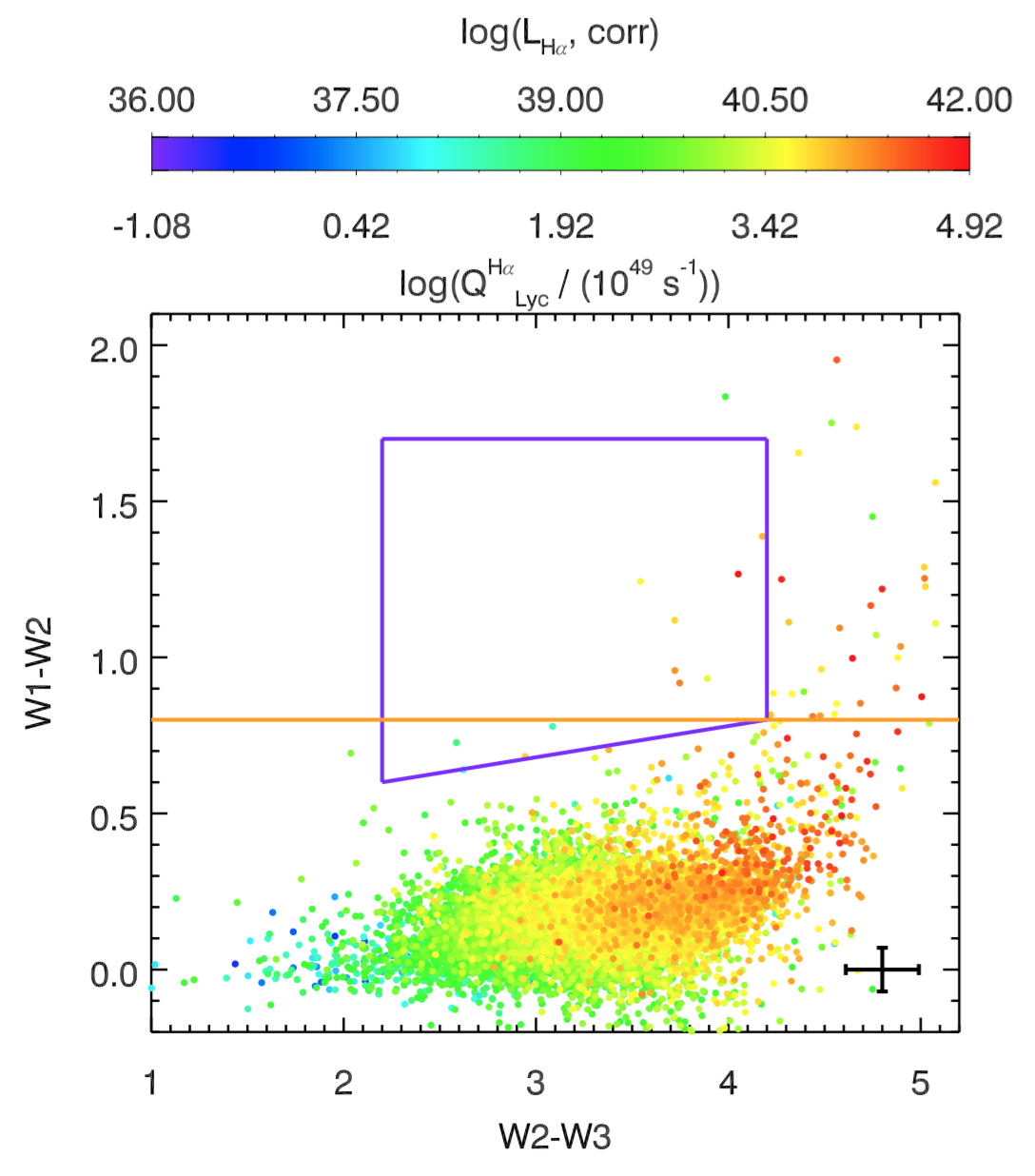} \\
  \end{tabular}
  \caption{\label{fig:wisecolorcolorhalpha} (Left) \textit{WISE} color-color diagram for the star-forming dwarf galaxies in our sample, colored by the EW of the H$\alpha$ emission line. The color bar indicates the derived ages of the stellar populations estimated from the H$\alpha$ EW and Starburst99 models. (Right) \textit{WISE} color-color diagram for the star-forming objects in our sample, colored by the extinction-corrected luminosity of the H$\alpha$ emission line. The color bar also provides an estimate for the rate of ionizing photons ($Q_{\mathrm{Lyc}}$) for these objects, as described in Equation \ref{ionizingphotons}. We also plot the median $\mathrm{W1}-\mathrm{W2}$ and $\mathrm{W2}-\mathrm{W3}$ uncertainties with the error bars in the bottom-right corner of each panel.}
\end{figure*}

We show \textit{WISE} color-color diagrams for the dwarf star-forming galaxies in Figure \ref{fig:wisecolorcolorhalpha}. We color the points based on the EW (left panel) and luminosity (right panel) of the H$\alpha$ emission line. In both panels of Figure \ref{fig:wisecolorcolorhalpha}, we also plot the same AGN selection boxes shown in Figure \ref{fig:reinesagn}. We can see that the majority of the objects lie outside the \citet{jarrett2011} selection box. At the reddest $\mathrm{W2}-\mathrm{W3}$ colors, the star-forming sequence extends upwards to redder $\mathrm{W1}-\mathrm{W2}$ colors, wrapping around the \citet{jarrett2011} AGN selection box and above the \citet{stern2012} selection line. In the left panel of Figure \ref{fig:wisecolorcolorhalpha}, we see that those objects with the reddest \textit{WISE} infrared colors have the largest H$\alpha$ EW values and youngest stellar ages, also indicated in Figure \ref{fig:optical_infrared_color_plot}. Similarly, in the right panel of Figure \ref{fig:wisecolorcolorhalpha}, we can see an even stronger trend for the objects with the reddest \textit{WISE} colors to have the largest H$\alpha$ luminosities, and rate of ionizing photons. 

\begin{figure*}[!htbp]
\centering
  \begin{tabular}{@{}cc@{}}
    \includegraphics[width=.43\textwidth]{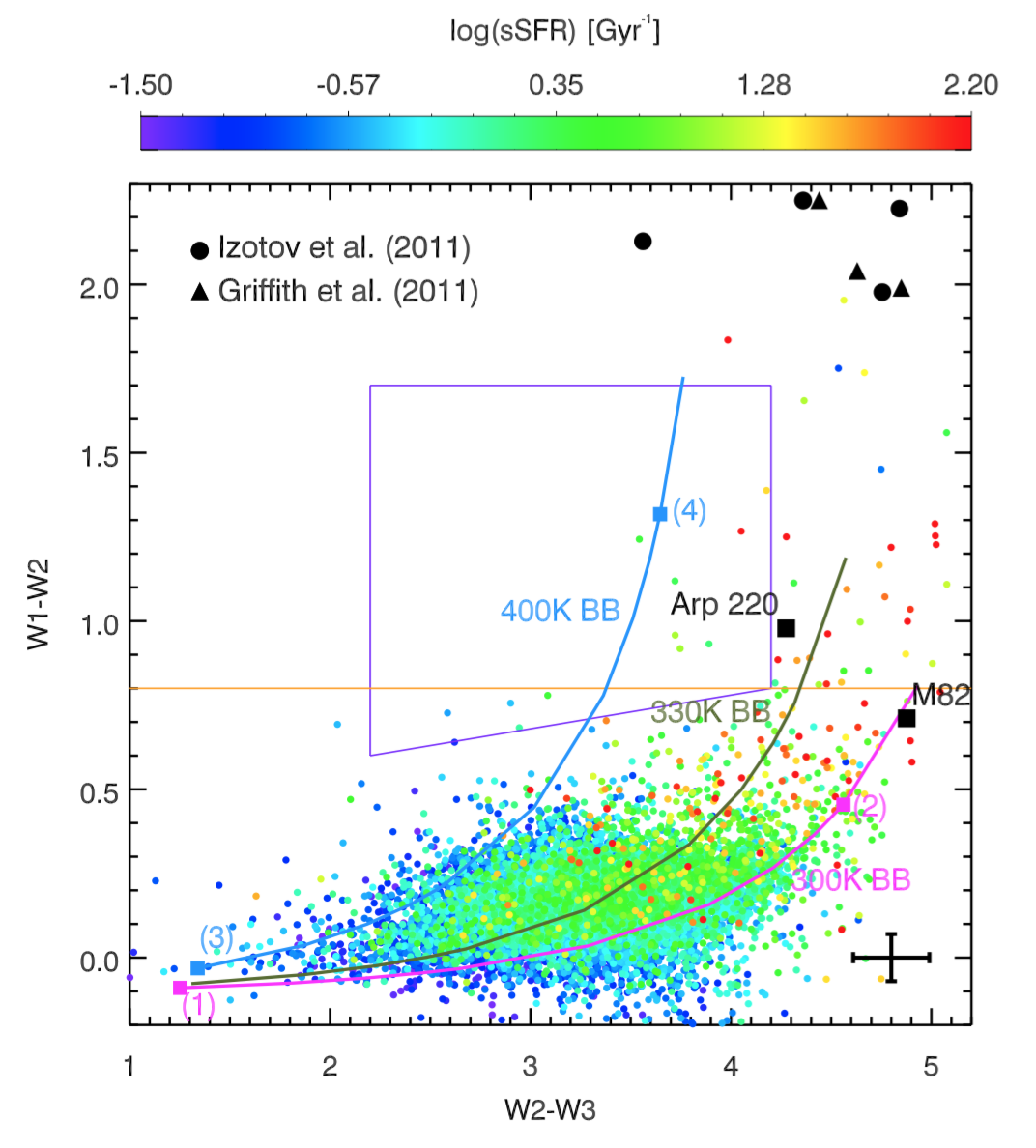} &
    \includegraphics[width=.53\textwidth]{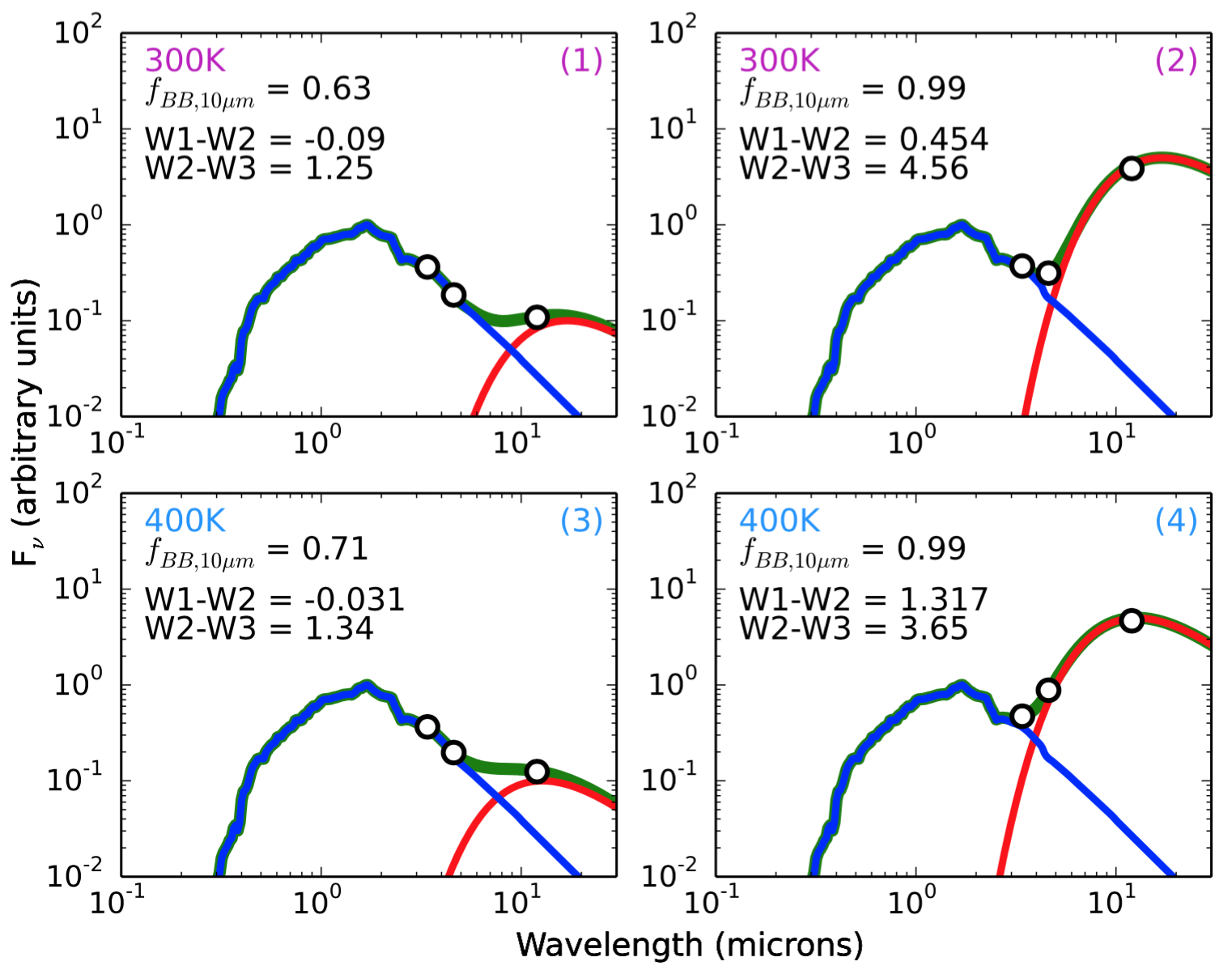} \\
  \end{tabular}
  \caption{\label{fig:wisecolorcolorwithtracks} \textit{WISE} color-color diagram for the BPT-selected star-forming dwarf galaxies. The objects are colored by sSFR as shown in the colorbar above the figure. Those objects with the highest sSFR values are found to the right and upwards on the diagram. We also plot two samples of objects with extreme \textit{WISE} $\mathrm{W1}-\mathrm{W2}$ colors from \citet{izotov2011} and \citet{griffith2011}, which appear to be a natural extension of the sequence of dwarf galaxies that extends from the bottom left corner to the top right corner. On top of the sequence, we plot color tracks generated by adding a blackbody to an \citet{assef2010} elliptical galaxy template (an elliptical galaxy template was chosen as the IR emission in this template is lacking in dust emission which we model with the blackbody) at $z = 0$. The different colors (magenta is 300 K, olive is 330 K, and blue is 400 K) represent different blackbody temperatures, and the sequence is generated by varying the ratio of the total IR luminosity of the black body with respect to the elliptical template, as shown in the sample SEDs plotted on the right panels. The colored numbers in the top right of each panel correspond to the positions on the left color-color diagram. In each panel on the right we also provide $f_{BB, 10\mu m}$, the fraction of the total flux that is emitted by the BB at 10$\mu$m. The dwarf galaxies seem to trace a sequence in color space represented by more luminous dust emission, with a scatter in the sequence that arises due to different maximum dust temperature. We also plot the median $\mathrm{W1}-\mathrm{W2}$ and $\mathrm{W2}-\mathrm{W3}$ uncertainties with the error bars in the bottom-right corner of the left panel.}
\end{figure*}

We explore the dwarf galaxy sequence on the \textit{WISE} color-color diagram as a function of dust temperature in Figure \ref{fig:wisecolorcolorwithtracks}. Here, on the left, we plot the same dwarf galaxies and AGN selection boxes as in Figure \ref{fig:wisecolorcolorhalpha}, but now the objects are colored by their specific star formation rates (sSFR). We see that the sequence of BPT-selected star-forming dwarf galaxies drives to the right and upward, and is correlated with the sSFR. We also plot extreme \textit{WISE} dwarf galaxies from \citet{griffith2011} and \citet{izotov2011}, which have $\mathrm{W1}-\mathrm{W2} \sim 2.0$. These rare objects are at the extreme end of the sequence traced by the star-forming dwarf galaxies in our sample. We also plot the \textit{WISE} colors for the infrared-bright star-forming galaxies Arp220 and M82, as measured from their SWIRE SED templates \citep{polletta2007}. Finally, in this panel we plot color tracks that we generated by linearly combining an elliptical galaxy template from \citet{assef2010} with a simple single-temperature blackbody (BB) at $z = 0$. For these tracks, we chose an elliptical galaxy template as this template does not show evidence for dust in the infrared. We plot three tracks at three different dust temperatures: 400 K (blue), 330 K (olive), and 300 K (magenta), and the position along the track (to the right and upwards) reflects the relative normalization of the BB and the galaxy template, which we parameterize using $f_{BB, 10 \mu m}$, the fraction of the total flux emitted from the blackbody at 10$\mu$m. In the right panels we show example SEDs for both the 300 K (top) and 400 K (bottom) examples, with the galaxy template in blue, the BB in red, and the total flux in green. We overplot the W1, W2, and W3 fluxes, and provide the $\mathrm{W1}-\mathrm{W2}$ and $\mathrm{W2}-\mathrm{W3}$ colors in each panel. 

From the tracks and SEDs in Figure \ref{fig:wisecolorcolorwithtracks}, we can see how the sequence of dwarf galaxies arises in \textit{WISE} color space. At a fixed BB normalization, as the temperature of the hottest dust in a given galaxy increases, galaxies move upward and to the left, satisfying the \citet{stern2012}, and potentially the \citet{jarrett2011} AGN selection criteria. Alternately, at a fixed BB temperature, as the normalization of the BB template increases, objects move upward and to the right. While a BB is a simplistic approximation of the dust emission observed in the infrared, our model does replicate the observed range in $\mathrm{W1}-\mathrm{W2}$ and $\mathrm{W2}-\mathrm{W3}$ colors for the dwarf galaxies. We emphasize that our model lacks PAH emission, which may significantly change the observed colors, although lower PAH abundance has been found in low-metallicity, high-sSFR dwarf galaxies \citep{wu2005, madden2006, engelbracht2008, galliano2008, remyruyer2015}. The BB temperatures from our models are consistent with those estimated for dust heated from star-formation in dwarf galaxies in the literature \citep{johnson2004, engelbracht2005,jackson2006,reines2008}. In addition, as the \textit{WISE} wavelength coverage does not extend to the far-infrared, and we cannot quantify cooler dust in these galaxies. We further discuss the implications of these results in Section \ref{sec:conclusions}.

\subsubsection{Mid-IR selected AGN Candidates}
\label{sec:AGNcandidates}

\begin{figure*}[!htbp]
	\centering
	  \begin{tabular}{@{}c@{}c@{}c@{}}
	   \includegraphics[width=.33\textwidth]{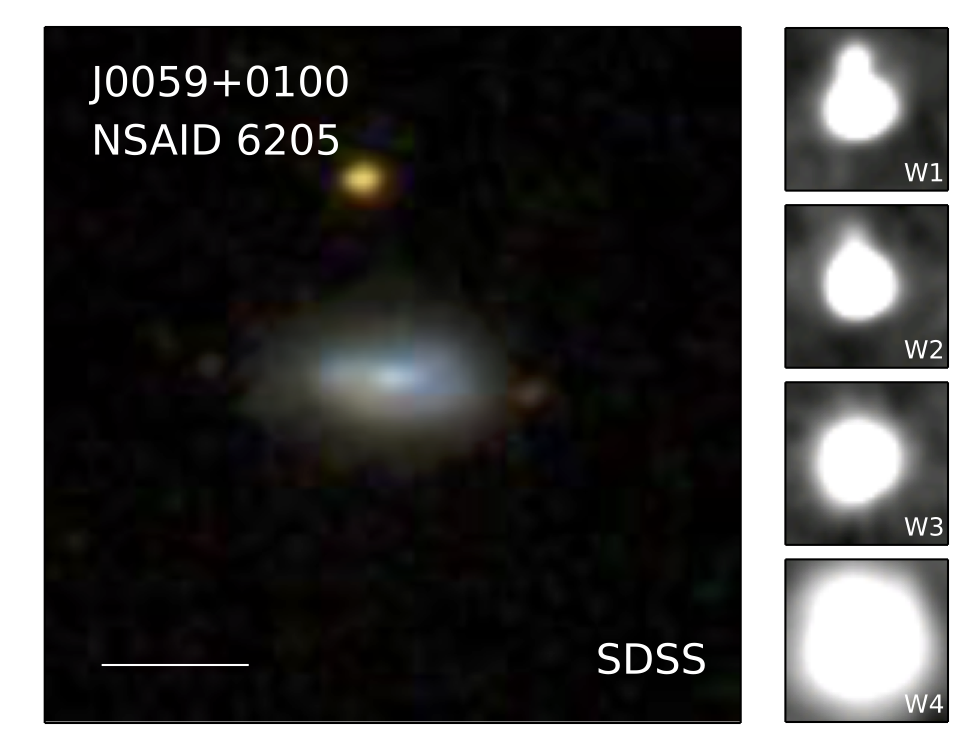}
	   \includegraphics[width=.33\textwidth]{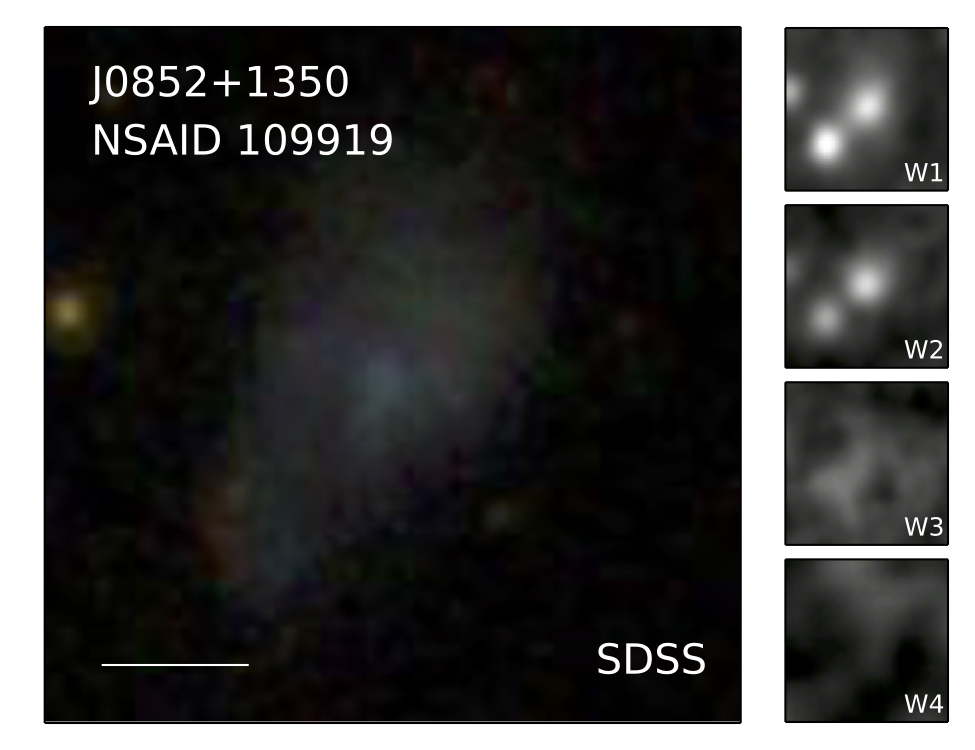} 
	   \includegraphics[width=.33\textwidth]{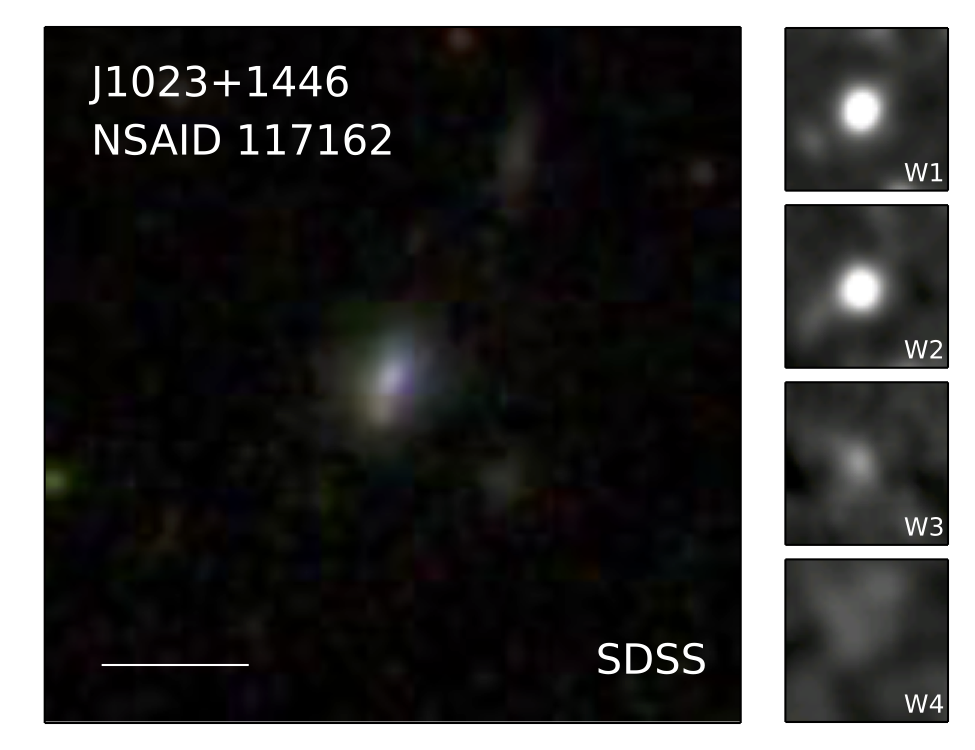} \\

	   \includegraphics[width=.33\textwidth]{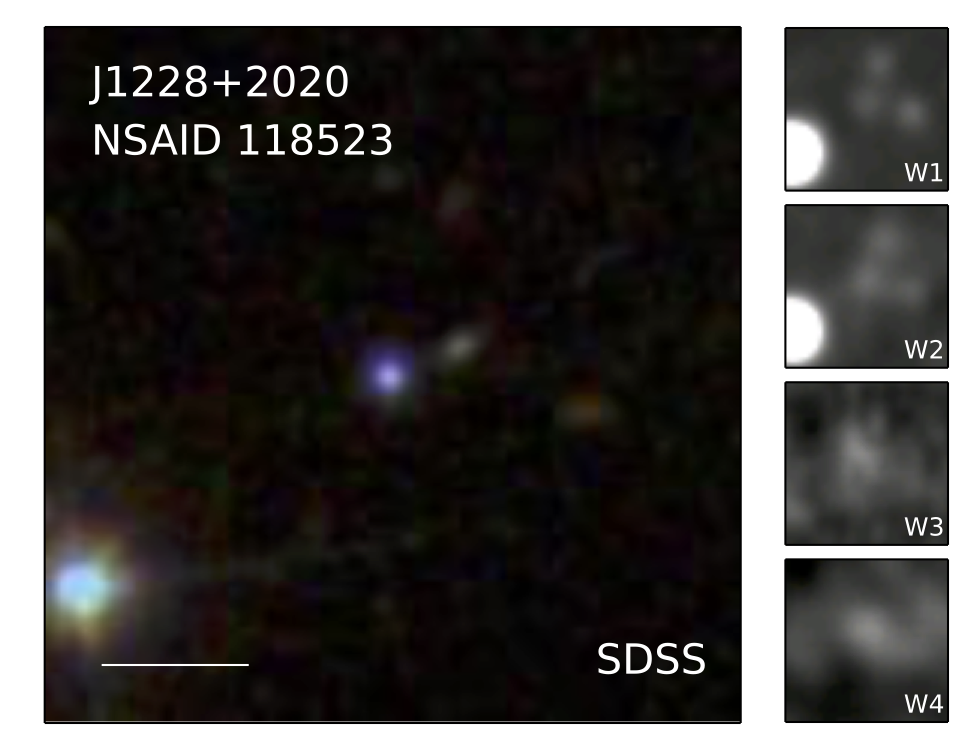} 
	   \includegraphics[width=.33\textwidth]{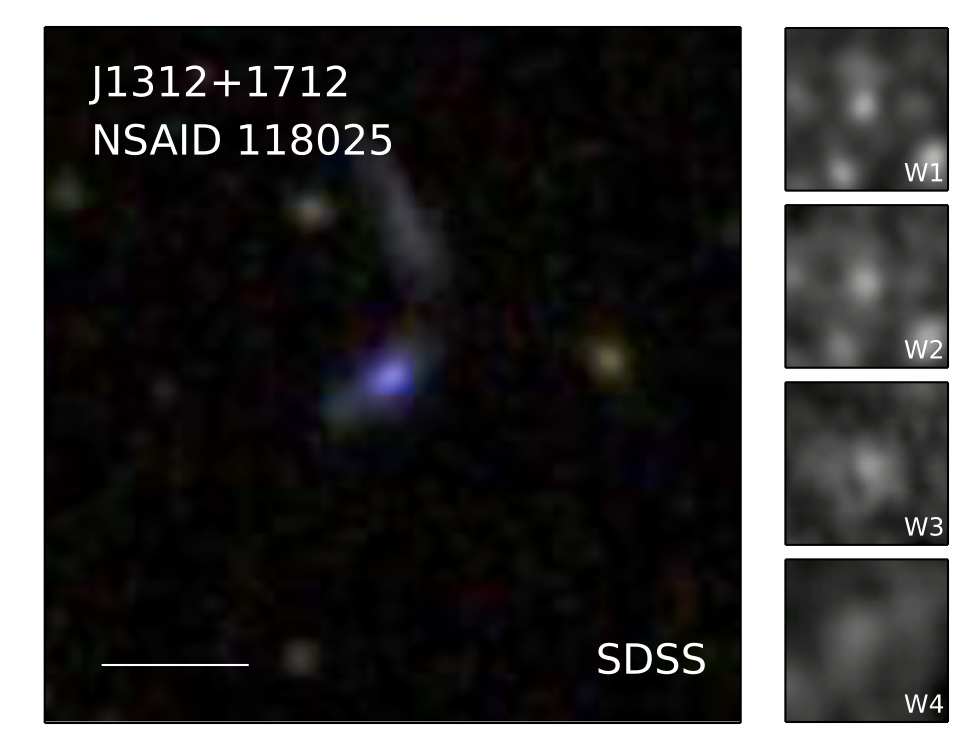}
	   \includegraphics[width=.33\textwidth]{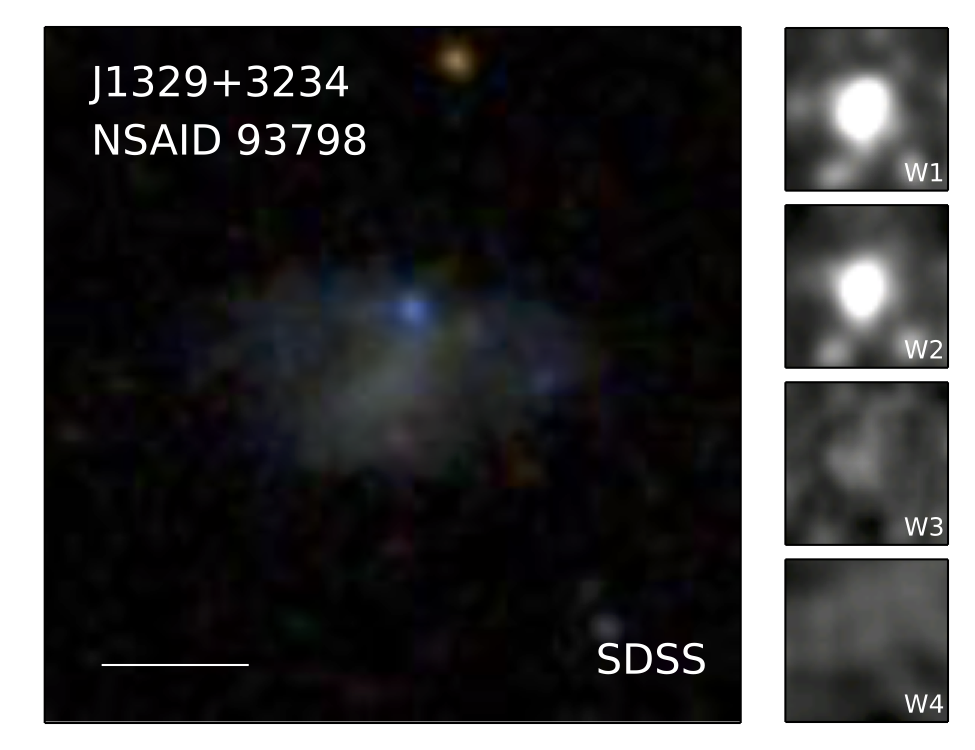} \\

	   \includegraphics[width=.33\textwidth]{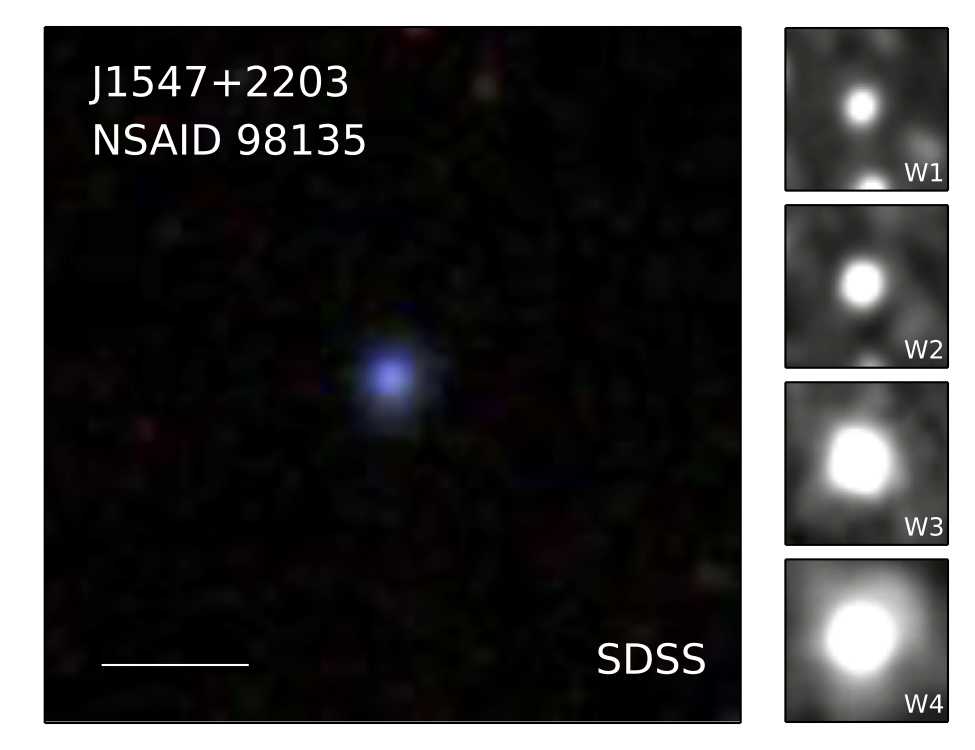}
	   \includegraphics[width=.33\textwidth]{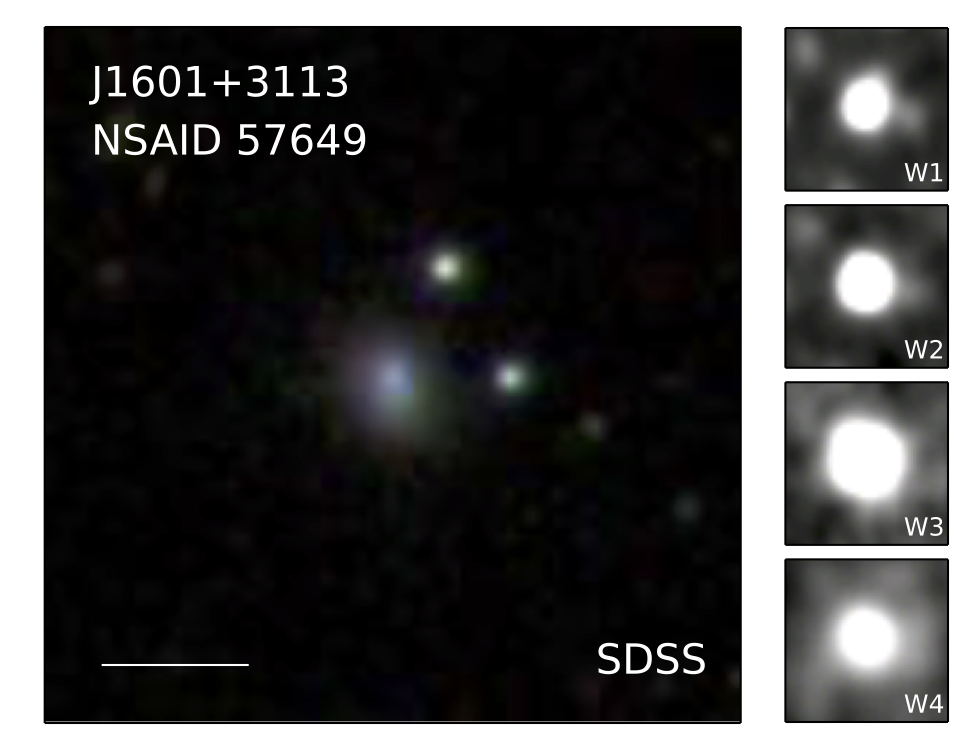}
	   \includegraphics[width=.33\textwidth]{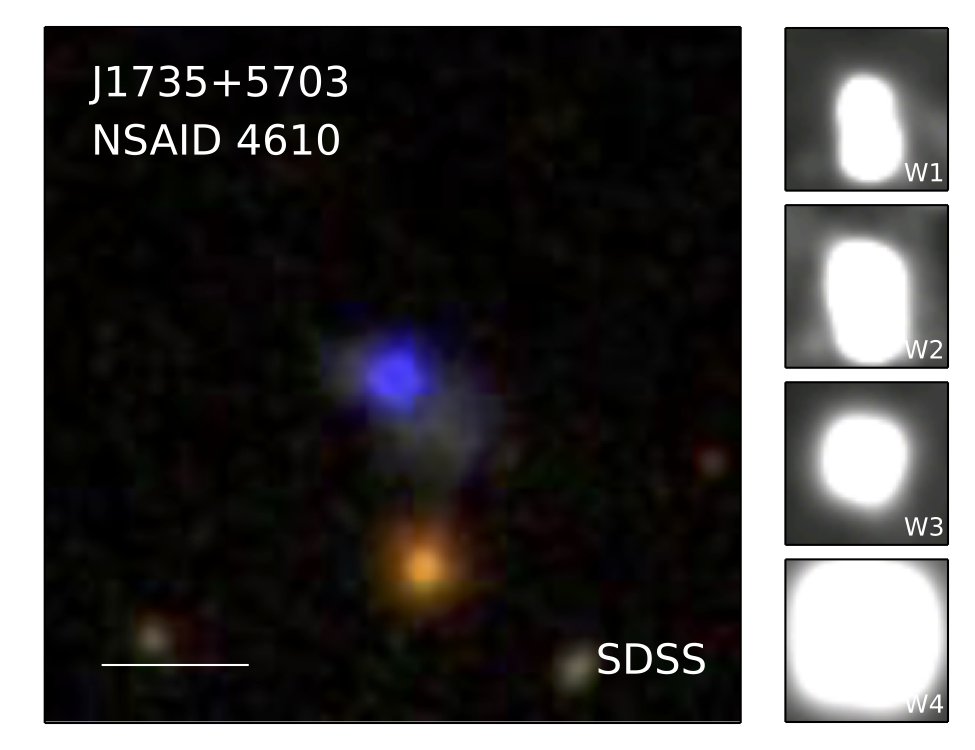} \\
	   	   
	   \includegraphics[width=.33\textwidth]{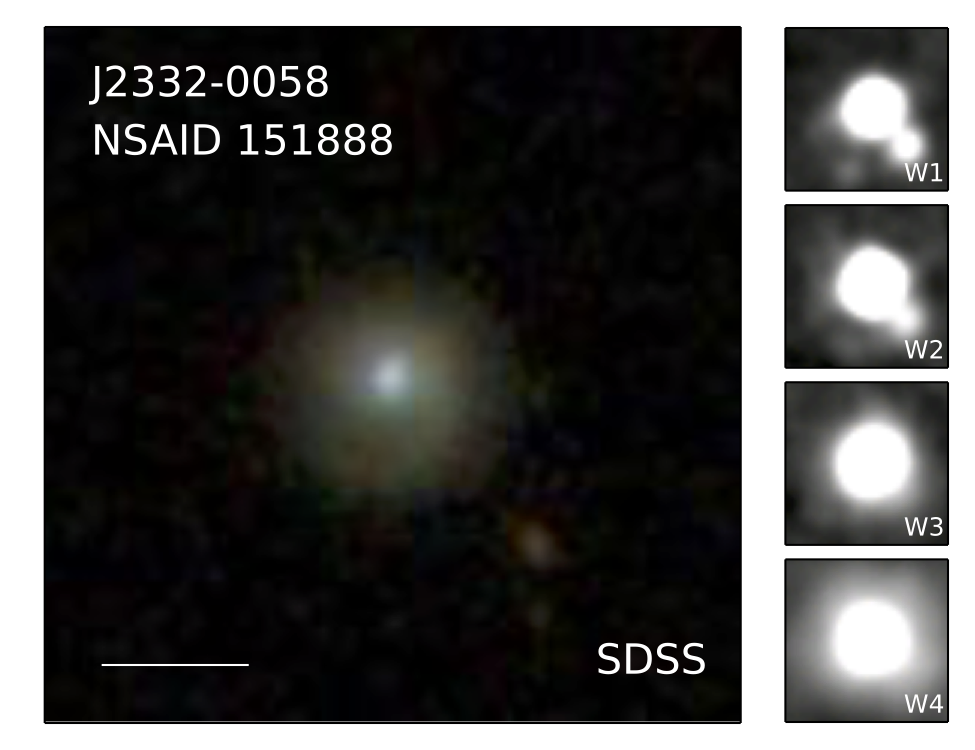}    

	\end{tabular}
	\caption{
	\label{fig:jarrett_thumbnails} SDSS and \textit{WISE} thumbnails for the \textit{WISE}-selected \citep{jarrett2011} AGN candidates in the star-forming region of the BPT diagram. In each panel, north is up, and east is to the left, and the bar in the bottom left of each SDSS panel is $10\arcsec$ in length. Both the SDSS and \textit{WISE} thumbnails are 48$\arcsec$ on a side. At the median redshift of our sample, $z = 0.03$, $1\arcsec = 0.6$ kpc.} 
         \end{figure*}

While our results indicate that the majority of star-forming dwarf galaxies lie outside of the commonly-used \textit{WISE} AGN selection boxes, there are potential IR-selected AGN candidates that we want to highlight. We find only 10 dwarf star-forming galaxies with \textit{WISE} colors that put them in the \citet{jarrett2011} AGN selection box. We present SDSS thumbnails for these objects in Figure \ref{fig:jarrett_thumbnails}. These objects span a range in optical color, but are on average bluer than the full sample of dwarf galaxies (e.g., see Figure \ref{fig:optical_infrared_color_plot}). Three out of the ten objects do not have significant (SNR $> 3$) W4 photometry (NSAID 93798, 109919, and 117162), as would be expected for an AGN. Two of these objects, NSAID 109919 and 93798, have low surface brightness, and do not show strong photometric evidence for a central nuclear source. NSA 93978 was presented by \citet{secrest2015} who demonstrated that this galaxy was detected by \textit{XMM-Newton} with a luminosity of $L_{2-10 \mathrm{keV}} = 2.4 \times 10^{40}$ erg s$^{-1}$. We further discuss this object in Section \ref{sec:comparison}. For the remainder of the objects, their morphologies are either compact blue nuggets (i.e. 98135), objects with blue cores and extended features (4610, and 118025), or objects with evidence for disks and nuclear cores (6205, 57649, and 151888). 

\subsection{Galaxies Without Optical Emission Line Flux Measurements}
\label{sec:noopticalemission}

There are a number of dwarf galaxies in our sample (4,469 objects) with low S/N detections of the emission lines in existing SDSS spectra or no spectroscopic information in the NSA. From this subsample, we initially found 68 objects that fall into the \citet{jarrett2011} selection box. We examined these objects to explore whether they had any alternate spectroscopic information: many have SDSS DR12 spectra which allowed us to determine that the NSA redshifts (and therefore masses) were incorrect for these objects, and they are likely unobscured quasars at $z \sim 1-2$. There are two objects, NSA IDs 110100 and 172349, which appear to be a chance alignment of a dwarf galaxy and a background quasar. After removing any quasar contaminants, we are left with only 25 potential IR-selected AGN dwarf galaxy candidates, and we show their SDSS and \textit{WISE} thumbnails in Figures \ref{fig:nonsdss_thumbnails}, \ref{fig:nonsdss_thumbnails_2}, and \ref{fig:nonsdss_thumbnails_3} in the Appendix. We note that two of these objects, NSA IDs 135305 and 150927, have poor W3 photometric fits (for these objects, the \textit{WISE} w3rchi2 value, the $\chi^2$ of the W3 profile fitting, is W3rchi2 $\geq 2$), and their W3 flux might be underpredicted, leading to these objects being bluer in $\mathrm{W2}-\mathrm{W3}$ color. Many of the dwarf galaxies do not have significant W4 detections. We would expect W4 detections from the power-law emission typically associated with an AGN, as seen for the BPT-AGNs in Figure \ref{fig:bpt_agns_thumbnails}. While the sizes and $g-r$ colors for these objects span a similar range as what was measured for the full sample of BPT star-forming objects in Figures \ref{fig:optical_infrared_color_plot} and \ref{fig:wisecolormorpho}, many are either optically redder or physically larger than the star-forming dwarf galaxies with red $\mathrm{W1}-\mathrm{W2}$ colors. Optical spectroscopy is necessary to better understand these systems. 

\begin{figure}[!htbp]
	\epsscale{1.2} 
	\plotone{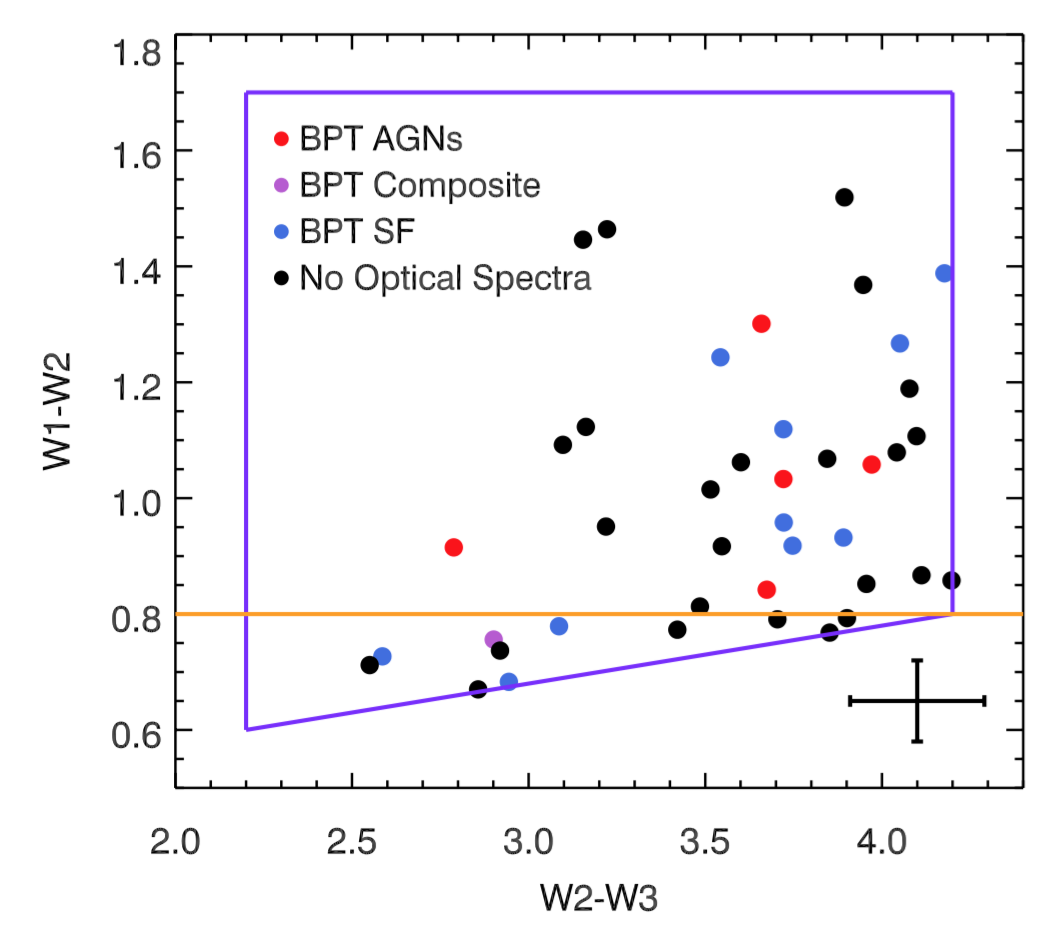}
	\caption{
	\label{fig:jarrettobjects} \textit{WISE} color-color diagram focusing on those objects that fall in the \citet{jarrett2011} selection box. We plot the positions of the \citet{reines2013} dwarf AGNs in red, dwarf composite objects in purple, the BPT SF objects in blue, and the galaxies without SDSS spectroscopic information in black. We also plot the median $\mathrm{W1}-\mathrm{W2}$ and $\mathrm{W2}-\mathrm{W3}$ uncertainties with the error bars in the bottom-right corner.} 
	\epsscale{1.}
         \end{figure}

We list the NSAID, SDSS ID, redshift, optical color, and \textit{WISE} colors for all of the AGN candidates (using the \citet{jarrett2011} selection box) in Table \ref{jarrettcandidates}. Furthermore, we plot all the NSA dwarf galaxies that have \textit{WISE} colors placing them inside the \citet{jarrett2011} selection box in Figure \ref{fig:jarrettobjects}, with the points colored by their optical properties. We note that these objects do not span the entire \citet{jarrett2011} selection space, but cluster near the right and bottom of the box. 

\subsection{Comparison to Recent Searches for AGNs in Dwarf Galaxies using Mid-IR Colors}
\label{sec:comparison}

Recently, \citet{satyapal2014} and \citet{sartori2015} used \textit{WISE} data to search for the presence of AGNs in dwarf and bulgeless galaxies. These objects were used to draw conclusions about the AGN population that would be missed in optical and X-ray surveys. Of the 30 AGN candidates using the \citet{jarrett2011} selection box and presented in \citet{satyapal2014}, only one object is in the mass and redshift range probed by our analysis here (the others are either at higher masses or higher redshifts), and it is found in our sample of star-forming dwarfs in the \citet{jarrett2011} box (NSA ID 93798). \citet{secrest2015} measured an X-ray luminosity from \textit{XMM-Newton} observations (extracted with a $\sim15\arcsec$ aperture) of $L_{2-10 \mathrm{keV}} = 2.4 \times 10^{40}$ erg s$^{-1}$ for this object, and claim that the $\mathrm{W1}-\mathrm{W2}$ color is strongly indicative of AGN activity. While this object is indeed in the \citet{jarrett2011} selection box, given the median uncertainties on the \textit{WISE} photometry, its infrared colors are consistent with this object potentially being scattered from the primary locus of star-forming galaxies and it does not have strong power law to longer infrared wavelengths as expected for an AGN.

\citet{sartori2015} presented a sample of 83 objects with photometry placing them in the \citet{jarrett2011} selection box, using data from the older \textit{WISE} All-Sky data release. These authors only required SNR $> 2$ for detections in the W1, W2, and W3 photometric bands. We cross-matched against the more up-to-date AllWISE selection, and using a SNR cut of 3.0, we find that only 34 objects ($\sim41$\%) continue to fall inside the \citeauthor{jarrett2011} selection box\footnote{All 83 of the objects in the \citet{sartori2015} sample have updated AllWISE W1 and W2 photometry with SNR $> 3.0$, but 26 of the objects have W3 photometry with SNR $< 3$, only 10 of which have $2 <$ SNR $< 3$.}. Of these 34 objects, only 15 are in our parent sample of dwarf galaxies. The remaining 19 objects are either not in the NSA because their redshifts are greater than 0.055, or are in the NSA but not classified as dwarf galaxies by our mass criterion. Looking closer at these 15 dwarf galaxies, 10 are found in Table \ref{jarrettcandidates} (NSA IDs 4610, 6205, 10311, 53934, 57649, 93798, 109919, 117162, 118025, and 118523), four of the 15 are BPT-selected AGNs or composite galaxies (NSA IDs 10779, 79874, 124050, and 125318), and one is the previously discussed quasar contaminant (NSA ID 110100). One object that was listed as a \citet{stern2012} candidate but not a \citet{jarrett2011} candidate in \citet{sartori2015} moved into the \citet{jarrett2011} selection box with the updated AllWISE photometry (NSA ID 98135), and one object we have listed in Table \ref{jarrettcandidates} does not appear in \citet{sartori2015} (NSA ID 151888). 

\subsection{The Feasibility of Detecting Supermassive Black Holes in Dwarf Galaxies with \textit{WISE}}
\label{sec:detection}

Finding evidence for AGNs in dwarf galaxies is challenging as AGNs at a given accretion rate have lower bolometric luminosities due to the smaller masses of the central black holes, making them harder to observe above the emission due to star formation in their host galaxies. It has been shown using common emission line AGN selection techniques that there is a strong bias against finding AGNs in low mass galaxies for this reason \citep{aird2012, bongiorno2012, hainline2012, trump2015}. To that end, we sought to explore the range of black hole masses in potential AGNs in dwarf galaxies that are detectable using \textit{WISE} colors, following a similar discussion in \citet{reines2013}. 

We calculate the minimum black hole detectable in the scenario where the AGN flux must be above some median flux from star formation at 3-4 $\mu$m. Assuming that the bulk of the objects with $\mathrm{W1}-\mathrm{W2} < 0.5$ have \textit{WISE} fluxes that are dominated by dust heated from star formation, we calculate the median W1 and W2 fluxes for all of these objects after shifting them to the median redshift of our sample. We do not perform this calculation using the W3 fluxes, as the differences in cold dust temperatures across the sample is more obvious at 12 $\mu$m than at the shorter wavelengths. We calculate median W1 and W2 fluxes of 382 and 243 $\mu$Jy respectively. From here, we use the low resolution AGN spectral template from \citet{assef2010} to estimate the AGN bolometric luminosity for an object at the median redshift of $z \sim 0.03$ given these flux limits. We note that this template was derived using more luminous AGNs, and the actual relationship between the IR and bolometric luminosities may vary for lower-mass black holes. We derive minimum black hole masses of $1-3\times10^3 \; \mathrm{M}_{\sun}$, if we assume that the black hole is accreting at the Eddington limit, where $L_{\mathrm{Edd}} = 1.3 \times 10^{38}(\mathrm{M}_{\mathrm{BH}}/\mathrm{M}_{\sun})$ erg s$^{-1}$. However, black hole accretion at such high levels is quite rare \citep{schulze2010}, and we also calculate that a black hole with mass $10^5 \; \mathrm{M}_{\sun}$ \citep[similar to the masses of the BHs measured in dwarf galaxies in ][]{reines2013,baldassare2015} would only need to radiate at $1-6\%$ of its Eddington luminosity to be observed given the median flux from star formation in our sample.

\section{Discussion and Conclusions}
\label{sec:conclusions}

To better understand the birth and growth of supermassive black holes, it is fundamentally important to find evidence for AGN activity in low mass dwarf galaxies. Here, we looked at the mid-IR properties of a sample of dwarf galaxies at $z < 0.055$ to explore the use of \textit{WISE} colors to select for AGN activity. We used the most up-to-date AllWISE photometry and individually inspected each candidate AGN to select a sample that is significantly smaller than previous samples of IR-selected AGNs in dwarf galaxies. 

Our main conclusions are:
\begin{itemize}
  \item The majority of optically-selected AGNs in dwarf galaxies have IR colors that are dominated by their host galaxies. 
  \item Dwarf galaxies with the reddest \textit{WISE} colors are compact, blue galaxies with young stellar ages and high sSFRs. Dwarf galaxies with extreme star-formation are capable of heating dust to temperatures producing $\mathrm{W1}-\mathrm{W2} > 0.8$, \citep[e.g.][]{stern2012}, and this single color cut alone should not be used to select AGNs in dwarf galaxies.
  \item We provide a sample of 41 dwarf galaxies in the NSA which have \textit{WISE} colors in the \citet{jarrett2011} AGN selection box, 6 of which have optical spectroscopic evidence for an AGN \citep{reines2013}. While the majority of the objects in our sample have been included in previous samples of \textit{WISE}-selected dwarf galaxy AGN candidates, our sample is much smaller due to the updated \textit{WISE} photometry, more conservative selection criteria and SNR thresholds, and the removal of spurious candidates. We caution that follow-up observations are necessary to confirm the presence of active massive black holes in the other 35 objects.
\end{itemize}

From our analysis, optically blue, high sSFR dwarfs with young starbursts and associated high ionizing fluxes can have red mid-IR colors that could be mistaken for AGN activity, particularly if using a simple $\mathrm{W1}-\mathrm{W2}$ color cut. While these objects have a population of young stars that can heat dust to temperatures which result in red $\mathrm{W1}-\mathrm{W2}$ colors ($\mathrm{W1}-\mathrm{W2} > 0.5-0.6$), they primarily have \textit{very} red $\mathrm{W2}-\mathrm{W3}$ colors ($\mathrm{W2}-\mathrm{W3} > 4.2$), which is less extreme, but similar to the $\mathrm{W2}-\mathrm{W3}$ colors seen for the dwarf star-forming galaxies from \citet{izotov2011}. 

AGNs, which can heat dust to even higher temperatures, are predominantly found in a different region of \textit{WISE} color-color space, and we only find 10 optically-selected star-forming dwarf galaxies that would be classified as an AGN by the \citet{jarrett2011} selection criteria. Of those objects, only 5 have strong W3 and W4 fluxes as would be expected for an AGN. In addition, these objects are consistent with the overall trend of star-forming galaxies in \textit{WISE} color space, and these objects could represent an extreme star-forming population that has scattered into the \citet{jarrett2011} selection box. We also found 25 additional dwarf galaxies that fell into the \citet{jarrett2011} box that do not have optical spectroscopy, or their SDSS emission lines were not strong enough to classify them on the BPT diagram. We include them as candidates, but we caution against using the mid-IR colors of these objects alone to classify them as AGNs; follow-up observations at optical or X-ray wavelengths would be helpful to understand the nature of their IR emission. 

Overall, it is important to use both $\mathrm{W1}-\mathrm{W2}$ and $\mathrm{W2}-\mathrm{W3}$ colors when selecting candidate AGNs in dwarf galaxies, as any selection of AGNs in dwarf galaxies that only uses $\mathrm{W1}-\mathrm{W2}$ color \citep[for instance, the selection method of][which has been demonstrated to be reliable at higher masses]{stern2012} will include a large amount of contamination from dwarf star-forming galaxies. Starting with the BPT star-forming galaxies discussed in Section \ref{sec:IRproperties}, we find that 42 dwarf galaxies have $\mathrm{W1}-\mathrm{W2} > 0.8$ (compared to only 10 in the \citet{jarrett2011} selection box), while 183 objects have $\mathrm{W1}-\mathrm{W2} > 0.5$. This contamination from star-forming dwarf galaxies is likely the cause of the observed rise in the fraction of AGN candidates at lower galaxy masses in \citet{satyapal2014} and \citet{sartori2015}. While we certainly cannot rule out the presence of AGNs in the optically-selected star-forming dwarf galaxies with red \textit{WISE} colors, the systematic correlations between star formation properties and infrared colors leads us to conclude that the infrared emission is unlikely to be powered by AGNs. Furthermore, the majority of known optically-selected AGNs in dwarf galaxies do not dominate the \textit{WISE} colors.

Our results are consistent with evidence in the literature that the ionizing UV radiation from young stars is one of the primary sources of dust heating in low-metallicity dwarf galaxies  \citep{izotov2014}. Both \textit{IRAS} and \textit{Spitzer} observations demonstrated that dwarf galaxies have evidence for large quantities of hot dust \citep{helou1986, hunter1989, melisseisrael1994, rosenberg2006, cannon2006}. In addition, it has been shown that the IR SED peak of dwarf galaxies is broader, which is often explained as resulting from dust at higher temperatures than what is observed in more massive galaxies \citep{boselli2012,smith2012,remyruyer2013,remyruyer2015, ciesla2014}. The temperatures from the tracks in Figure \ref{fig:wisecolorcolorwithtracks} for the hot dust that would be necessary to produce the observed \textit{WISE} colors has been observed in local low-metallicity galaxies from \textit{Spitzer} 8$\mu$m observations \citep{engelbracht2005,jackson2006}, and even hotter dust has been invoked to explain the near-IR excesses observed in the dwarf galaxies SBS 0335-052 \citep{reines2008} and Haro 3 \citep{johnson2004}. Recently, in an analysis of the infrared properties of a large sample of low-metallicity dwarf galaxies, \citet{remyruyer2015} demonstrated that the dust SED for these galaxies peaks at shorter wavelengths as compared to higher-metallicity systems, which they attribute to a clumpy interstellar medium that allows for a wider range of dust temperatures. In addition, the lower-metallicity dust will attenuate the light from young stars less, and the dust can then be heated deeper within individual molecular clouds. \citet{cormier2015} used \textit{Herschel} PACS spectroscopy of low-metallicity dwarf galaxies to provide evidence that the interstellar medium in these galaxies is more porous than in metal-rich galaxies, leading to a larger fraction of the stellar UV radiation heating dust. These results are supported by modeling by \citet{hirashita2004}, which found that dust temperature and dust luminosity is higher in dense, compact, low-metallicity star-forming regions. 

\acknowledgments 
Support for AER was provided by NASA through Hubble Fellowship grant HST-HF2-51347.001-A awarded by the Space Telescope Science Institute, which is operated by the Association of Universities for Research in Astronomy, Inc., for NASA, under contract NAS 5-26555. The work of DS was carried out at the Jet Propulsion Laboratory, California Institute of Technology, under a contract with NASA. We are grateful to the entire SDSS collaboration for providing the data that made this work possible, and to Michael Blanton and all those involved in creating the NASA-Sloan Atlas. This publication makes use of data products from the Wide-field Infrared Survey Explorer, which is a joint project of the University of California, Los Angeles, and the Jet Propulsion Laboratory/California Institute of Technology, and NEOWISE, which is a project of the Jet Propulsion Laboratory/California Institute of Technology. \textit{WISE} and NEOWISE are funded by the National Aeronautics and Space Administration.

\bibliographystyle{apj}

% TABLES AND FIGURES --------------------------------------

\begin{deluxetable}{lllrcrr}
\centering
\tabletypesize{\scriptsize}
\tablecaption{\textit{WISE}-selected / \citet{jarrett2011} AGN candidates \label{jarrettcandidates}}
\tablewidth{0pt}
\tablehead{
\colhead{NSAID} & \colhead{SDSS ID} & \colhead{$z$} & \colhead{$g-r$} & \colhead{W1} & \colhead{W1-W2} & \colhead{W2-W3}
}
\startdata
\cutinhead{BPT-selected AGNs$^{a}$}
\\
68765 & J032224.64+401119.8 & 0.026 & 0.618 & 14.012 $\pm$ 0.020 & 0.842 & 3.674 \\ 
10779 & J090613.75+561015.5 & 0.047 & 0.426 & 13.400 $\pm$ 0.024 & 1.301 & 3.659 \\ 
125318 & J095418.15+471725.1 & 0.033 & 0.448 & 13.859 $\pm$ 0.027 & 1.058 & 3.971 \\ 
113566 & J114359.58+244251.7 & 0.050 & 0.683 & 14.275 $\pm$ 0.028 & 0.915 & 2.788 \\ 
104527 & J133245.62+263449.3 &  0.047 & 0.281 & 13.514 $\pm$ 0.025 & 1.033 & 3.721 \\ 
\cutinhead{BPT-selected Composite Galaxies$^{a}$}
\\
79874 & J152637.36+065941.6 & 0.0384 & 0.299 & 14.121 $\pm$ 0.026 & 0.756 & 2.901 \\ 
\cutinhead{BPT-selected Star-forming Galaxies$^{b}$}
\\
6205 & J005904.10+010004.2 & 0.018 & 0.240 & 13.308 $\pm$ 0.024 & 0.932 & 3.891 \\  
109919 & J085233.75+135028.3 & 0.005 & 0.305 & 15.761 $\pm$ 0.052 & 0.779 & 3.086 \\ 
117162 & J102345.04+144604.7 & 0.047 & 0.218 & 14.794 $\pm$ 0.033 & 0.683 & 2.944 \\ 
118523 & J122822.81+202043.6 & 0.049 & $-$0.053 & 16.432 $\pm$ 0.074 & 0.918 & 3.747 \\ 
118025 & J131253.76+171231.1 & 0.052 & $-$0.260 & 16.583 $\pm$ 0.086 & 0.958 & 3.722 \\ 
93798 & J132932.42+323416.9 & 0.016 & 0.252 & 15.297 $\pm$ 0.036 & 0.727 & 2.586 \\ 
98135 & J154748.99+220303.2 & 0.031 & $-$0.390 & 15.583 $\pm$ 0.041 & 1.388 & 4.177 \\ 
57649 & J160135.95+311353.7 & 0.031 & 0.136 & 14.526 $\pm$ 0.028 & 1.243 & 3.543 \\ 
4610 & J173501.25+570308.8 & 0.047 & $-$1.003 & 13.041 $\pm$ 0.029 & 1.267 & 4.051 \\ 
151888 & J233244.60$-$005847.9 & 0.024 & 0.382 & 13.581 $\pm$ 0.011 & 1.119 & 3.721 \\ 
\cutinhead{Galaxies Without Optical Emission Line Fluxes$^{c}$}
 \\
127223 & J003441.39$-$212928.9 & 0.023 & 0.027 & 15.975 $\pm$ 0.055 & 0.793 & 3.901 \\ 
127548 & J004206.29$-$093335.0 & 0.054 & 0.515 & 15.659 $\pm$ 0.049 & 1.189 & 4.078 \\ 
128988 & J011600.59+063812.9 & 0.008 & 0.395 & 14.991 $\pm$ 0.035  & 0.737 & 2.919 \\ 
130346 & J014554.39+241555.0 & 0.035 & $-$0.467 & 16.165 $\pm$ 0.056 & 1.107 & 4.098 \\ 
130679 & J015446.09+014326.0 & 0.039 & 0.208 & 16.544 $\pm$ 0.074 & 0.813 & 3.485 \\ 
170240 & J023850.49+272159.0 & 0.005 & 0.062 & 14.688 $\pm$ 0.030 & 0.712 & 2.550 \\ 
64260 & J075824.77+202352.3 & 0.035 & $-$0.631 & 16.317 $\pm$ 0.074 & 1.519 & 3.894 \\ 
135305 & J083538.40$-$011407.0 & 0.044 & 0.812 & 12.038 $\pm$ 0.006 & 1.015 & 3.515 \\ 
10311 & J084624.06+540915.9 & 0.031 & 0.434 & 15.606 $\pm$ 0.042 & 0.951 & 3.219 \\ 
135720 & J091108.59+541051.9 & 0.027 & $-$0.596 & 15.816 $\pm$ 0.042 & 0.768 & 3.852 \\ 
135754 & J091305.00+284845.0 & 0.027 & $-$0.285 & 16.998 $\pm$ 0.119 & 1.079 & 4.042 \\ 
136250 & J095643.10$-$030409.0 & 0.047 & 0.700 &14.258 $\pm$ 0.028  & 0.670 & 2.857 \\ 
138274 & J105301.99+365811.9 & 0.026 & $-$0.268 & 16.522 $\pm$ 0.077 & 0.852 & 3.956 \\ 
141379 & J122655.39+424208.0 & 0.009 & 1.058 & 14.945 $\pm$ 0.032 & 0.858 & 4.197 \\ 
171667 & J123948.06$-$171753.9 & 0.028 & 0.788 & 13.990 $\pm$ 0.026 & 0.867 & 4.112 \\ 
142435 & J125037.90+434519.9 & 0.039 & 0.418 & 15.316 $\pm$ 0.036 & 1.464 & 3.222 \\ 
142555 & J125615.79$-$032723.0 & 0.045 & 0.146 & 16.044 $\pm$ 0.054 & 0.773 & 3.421 \\ 
143373 & J132119.70+320824.9 & 0.018 & $-$0.019 & 16.823 $\pm$ 0.088 & 1.068 & 3.845 \\ 
144502 & J140609.20$-$032224.0 & 0.055 & 0.267 & 15.951 $\pm$ 0.050 & 0.917 & 3.547 \\ 
53934$^d$ & J155320.20+420735.6 & 0.022 & 0.443 & 15.757 $\pm$ 0.037 & 1.092 & 3.097 \\ 
171765 & J204609.92$-$160058.2 & 0.033 & 0.240 & 14.964 $\pm$ 0.037 & 1.123 & 3.162 \\
44938$^d$ & J211012.75+002428.2 & 0.020 & 0.638 & 16.368 $\pm$ 0.076 & 1.446 & 3.154 \\ 
150295 & J225515.30+180835.0 & 0.043 & $-$0.192 & 13.158 $\pm$ 0.025  & 1.368 & 3.947 \\ 
150927 & J231544.60+065439.0 & 0.008 & 0.043 & 12.430 $\pm$ 0.007 & 0.791 & 3.704 \\
152731 & J235005.99+163505.9 & 0.049 & $-$0.575 & 16.725 $\pm$ 0.099 & 1.062 & 3.601 
\enddata
\tablenotetext{a}{Objects from \citet{reines2013}. The SDSS and \textit{WISE} thumbnails for these objects are shown in Figure \ref{fig:bpt_agns_thumbnails}.}
\tablenotetext{b}{The SDSS and \textit{WISE} thumbnails for these objects are shown in Figure \ref{fig:jarrett_thumbnails}.}
\tablenotetext{c}{The SDSS and \textit{WISE} thumbnails for these objects are shown in Figure \ref{fig:nonsdss_thumbnails}, \ref{fig:nonsdss_thumbnails_2}, and \ref{fig:nonsdss_thumbnails_3} in the Appendix.}
\tablenotetext{d}{These objects have SDSS spectra, but one or more of the the emission lines used for BPT selection lack sufficient S/N for our optical classification.}
\end{deluxetable}

\appendix 
\section{Thumbnails of Jarrett et al. IR-AGN Candidates Without Optical Emission Line Flux Measurements}
\label{sec:appendix}

	\begin{figure*}[htb]
	\centering
	  \begin{tabular}{@{}c@{}c@{}c@{}}
	   \includegraphics[width=.33\textwidth]{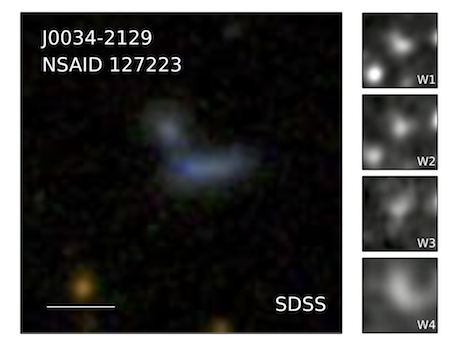}
	   \includegraphics[width=.33\textwidth]{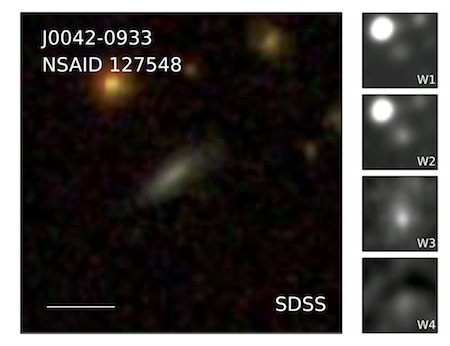}
	   \includegraphics[width=.33\textwidth]{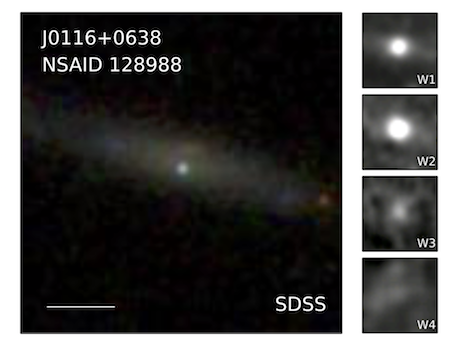} \\

	   \includegraphics[width=.33\textwidth]{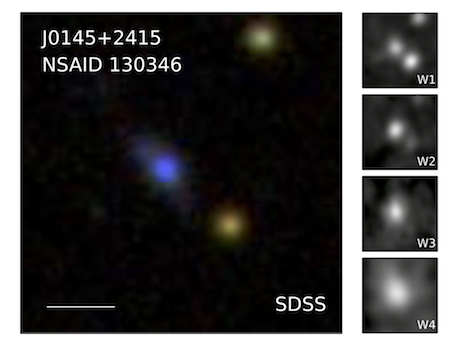}
	   \includegraphics[width=.33\textwidth]{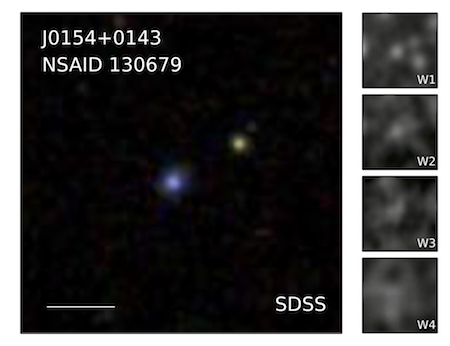}
	   \includegraphics[width=.33\textwidth]{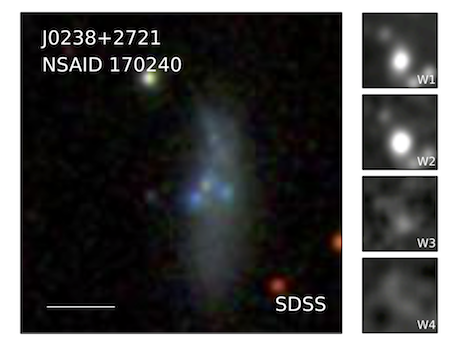} \\

	   \includegraphics[width=.33\textwidth]{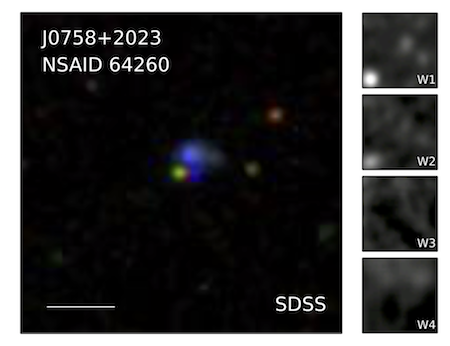}
	   \includegraphics[width=.33\textwidth]{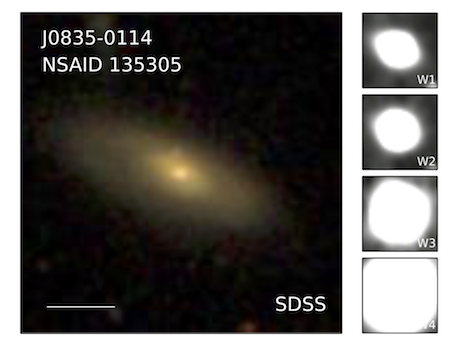}
	   \includegraphics[width=.33\textwidth]{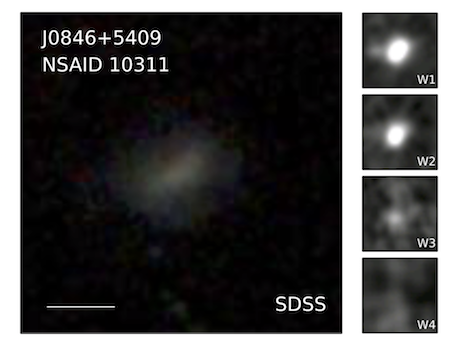} \\

	   \includegraphics[width=.33\textwidth]{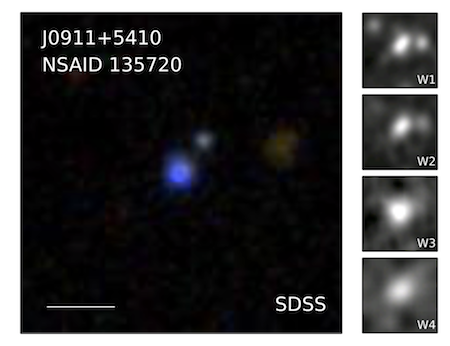}
	   \includegraphics[width=.33\textwidth]{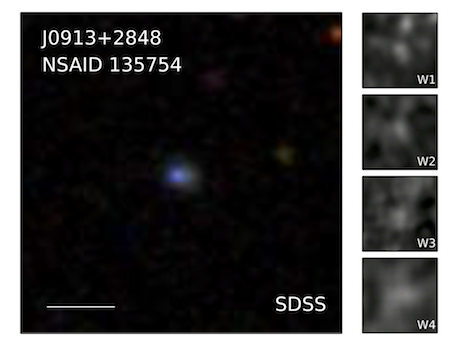}
	   \includegraphics[width=.33\textwidth]{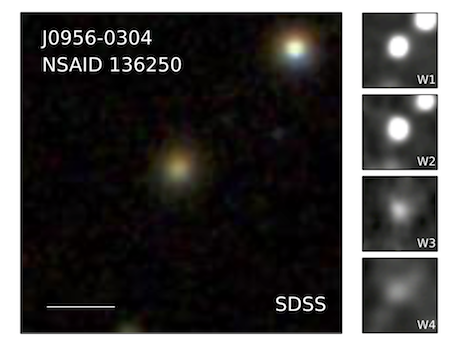} \\

	\end{tabular}
	\caption{
	\label{fig:nonsdss_thumbnails} SDSS and \textit{WISE} thumbnails for the Jarrett et al. \textit{WISE} AGN candidates without optical emission line flux measurements. Each object is labelled as in Figure \ref{fig:jarrett_thumbnails}.} 
         \end{figure*}

	\begin{figure*}[htb]
	\centering
	  \begin{tabular}{@{}c@{}c@{}c@{}}
	   \includegraphics[width=.33\textwidth]{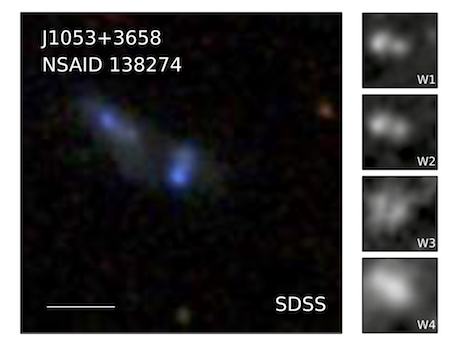}
	   \includegraphics[width=.33\textwidth]{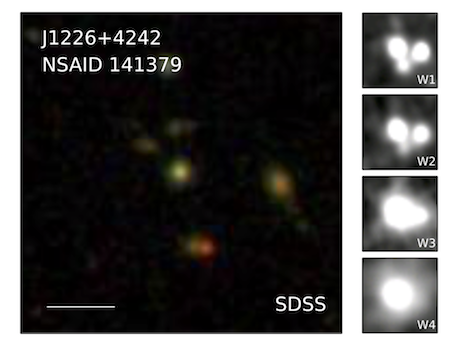}
	   \includegraphics[width=.33\textwidth]{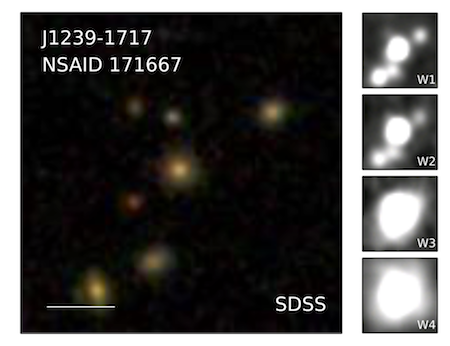} \\

	   \includegraphics[width=.33\textwidth]{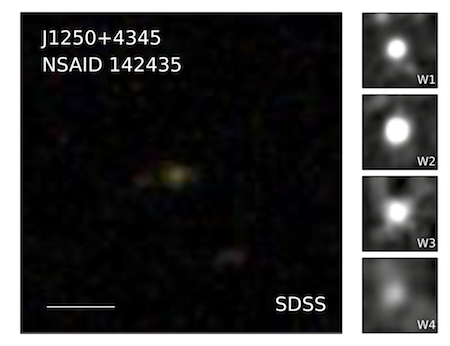}
	   \includegraphics[width=.33\textwidth]{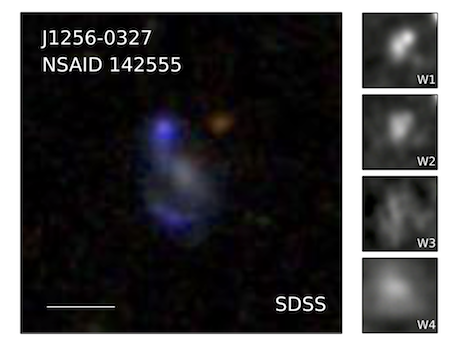}
	   \includegraphics[width=.33\textwidth]{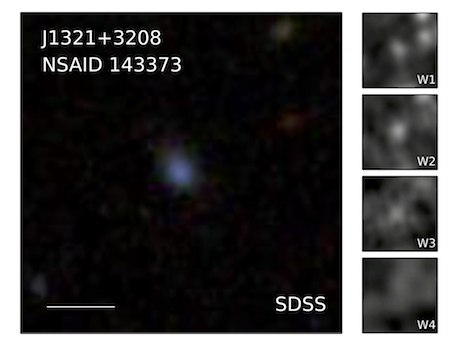} \\

	   \includegraphics[width=.33\textwidth]{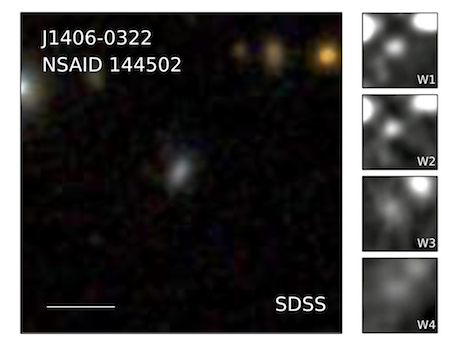}
	   \includegraphics[width=.33\textwidth]{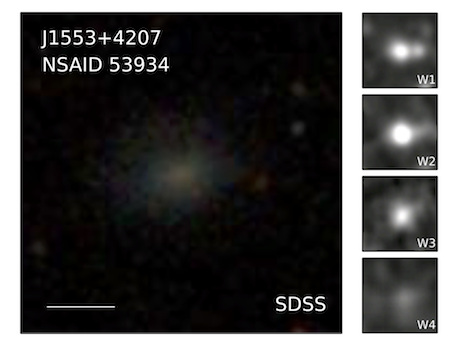}
	   \includegraphics[width=.33\textwidth]{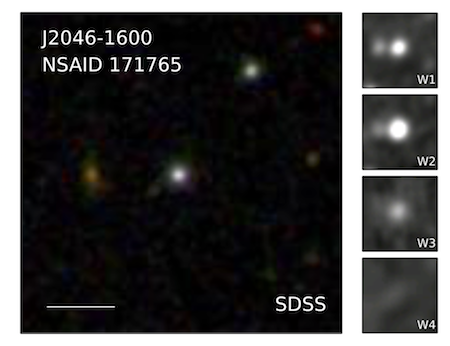} \\

	   \includegraphics[width=.33\textwidth]{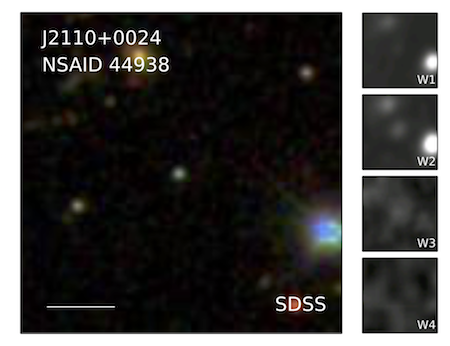}
	   \includegraphics[width=.33\textwidth]{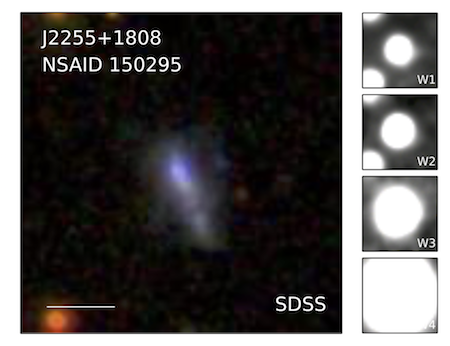}
	   \includegraphics[width=.33\textwidth]{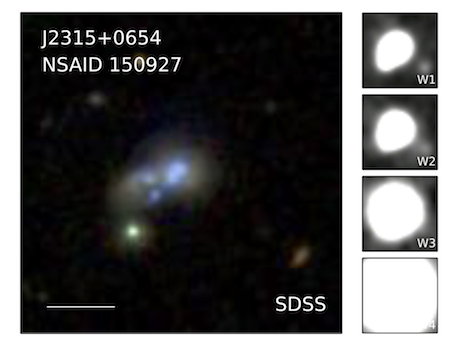} \\

	\end{tabular}
	\caption{
	\label{fig:nonsdss_thumbnails_2} (Continued from Figure \ref{fig:nonsdss_thumbnails}) SDSS and \textit{WISE} thumbnails for the Jarrett et al. \textit{WISE} AGN candidates without optical emission line flux measurements. Each object is labelled as in Figure \ref{fig:jarrett_thumbnails}.} 
         \end{figure*}

	\begin{figure*}[htb]
	\centering
	  \begin{tabular}{@{}c@{}c@{}c@{}}
	   \includegraphics[width=.33\textwidth]{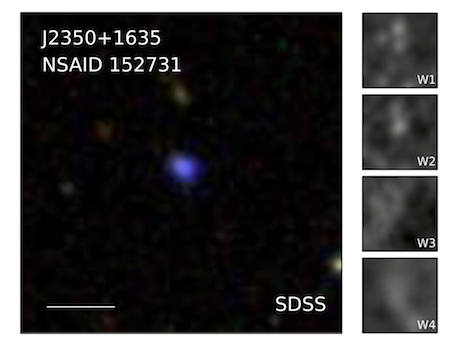} \\

	\end{tabular}
	\caption{
	\label{fig:nonsdss_thumbnails_3} (Continued from Figure \ref{fig:nonsdss_thumbnails}) SDSS and \textit{WISE} thumbnails for the Jarrett et al. \textit{WISE} AGN candidates without optical emission line flux measurements. Each object is labelled as in Figure \ref{fig:jarrett_thumbnails}.} 
         \end{figure*}
          
\end{document}